\documentclass[journal=jpc,manuscript=article]{achemso}
\usepackage[utf8]{inputenc}

\usepackage{longtable}
\usepackage{multirow}
\usepackage{amsmath}
\usepackage{subfigure}
\usepackage[usenames,dvipsnames]{xcolor}
\usepackage{soul}
\usepackage{bm}

\usepackage{amssymb}
\usepackage{siunitx}
\usepackage{multirow}
\usepackage{multicol} \usepackage{wrapfig} \usepackage{enumitem}
\usepackage{bm} \usepackage{outlines} 
\usepackage{url}
\usepackage{hyperref}
\usepackage{xr-hyper}

\usepackage{siunitx}\usepackage{booktabs} 
\usepackage{soul}

\usepackage[normalem]{ulem}
\makeatletter
\newcommand*{\addFileDependency}[1]{% argument=file name and extension
  \typeout{(#1)}
  \@addtofilelist{#1}
  \IfFileExists{#1}{}{\typeout{No file #1.}}
}
\makeatother

\SectionNumbersOn

\author{Kham Lek Chaton, Eric D. Boittier, Mike Devereux}
\author{Markus Meuwly} \affiliation[University of Basel]{Department of
  Chemistry, University of Basel, Klingelbergstrasse 80, CH-4056
  Basel, Switzerland.}  \email{m.meuwly@unibas.ch}

\title{Explicit, Machine-Learned Two-Body Potentials for Molecular
  Simulations}

\keywords{}

\begin{document}
\date{\today}

\begin{abstract}
A new pairwise hybrid machine-learning/molecular mechanics (ML/MM)
potential is introduced that is conceived for application to large,
heterogeneous condensed-phase systems. The PhysNet ML method describes
monomers and short-range dimer interactions, while a classical MM
force field describes pairwise interactions beyond a defined switching
distance. Models are fitted to MP2 dimer and pairwise cluster
energies, and the quality of each model is assessed at different
switching distances and using MM approaches with and without detailed
distributed charge electrostatics. The applicability of the approach
to molecular dynamics simulations is demonstrated for a basic
implementation applied to a small model system. Dichloromethane and
acetone are used as test systems to demonstrate the accuracy of the
approach in describing pairwise reference data, and also to highlight
the limitations of the pairwise approach for systems that exhibit
significant many-body effects in condensed phase, paving the way for
the addition of a general many-body correction in future work.\\
\end{abstract}

\vspace{0.5cm}
\noindent \textbf{\small Keywords: Machine Learning, Potential Energy
  Surface, Long Range Interactions}

\maketitle

\section{Introduction}
Empirical energy functions (EEFs) are a successful concept for
modeling chemical and biological materials and processes at various
scales. One of the hallmarks of a meaningful {\it model} is the fact
that it can be improved in a targeted fashion. For example, a model
can be improved by changing several or all parameters in view of
experiments. Depending on how this is done, a deeper understanding may
be obtained. This has, for example, been attempted by morphing
potential energy surfaces (PESs) given measurements of rovibrational
energies or kinetic energy distributions for small
molecules.\cite{MM.morph:99} In this case, the fitted model provides a
new, improved PES which can be used in new simulations. On the other
hand, biasing a given energy function in view of measured infrared
transitions and their relative intensities for cationic trialanine
yielded a Ramachandran map consistent with NMR experiments. However,
because only the RM was biased and not the underlying energy function,
no improved PES was available for running new simulations.\\

\noindent
Yet another approach for improving empirical force fields (EFFs) is to
declare each (additive) contribution as replaceable by a more
physics-motivated and/or computationally improved model and
representation. This was recently pursued for electrostatic
contributions and within the formulation of EEFs.\cite{MM.ffs:2024}
However, the advent of machine learning-based potential energy
surfaces (ML-PESs) has brought additional opportunities for targeted
improvement of models to represent inter- and intramolecular
interactions.\cite{unke:2021,MM.rev:2022} Such models come in very
different flavours and range from highly accurate ML-PESs for single
systems to foundational models for particular applications.\\

\noindent
Much recent progress has been made on computational approaches that
increase the accuracy of condensed phase simulations while maintaining
computational cost at a level where sufficiently large system sizes
and sufficiently long timescales for condensed phase properties to
converge remain accessible. They can be broadly divided into pure
machine learning-based (ML) methods, ``next-generation'' molecular
mechanics (MM), and hybrid ML/MM. ``Deep Potential Molecular
Dynamics'' (DPMD)\cite{zhang:2018} is one example of a pure ML
approach, where atomic energies are a function of their local
environment up to a cutoff, thereby including local many-body effects
but neglecting explicit long-range electrostatic interactions. Pure
``next generation'' MM approaches with increased accuracy due to more
comprehensive functional forms include SIBFA,\cite{Naseem-Khan:2022}
designed to describe each term of a reference {\it ab initio} energy
decomposition scheme in order to recover the total system energy, and
AMOEBA,\cite{ren:2003} another multipolar, fully polarizable classical
potential. Particularly promising are various hybrid ML/MM approaches,
such as the SO3LR\cite{kabylda:2025} model that combines the SO3krates
NN for semilocal interactions with a pairwise force field for
short-range repulsion, long-range electrostatics and dispersion
interactions. Also the QCT-FF\cite{popelier:2015} method that
partitions the wave function and with it the total system energy into
atomic contributions using the topology of the molecular electron
density. A NN then learns the atomic energy contributions from their
local environment, plus associated atomic multipole moments are used
to evaluate long-range electrostatics. Finally, the
MB-pol\cite{Bore:2023} model combines permutationally invariant
polynomials that describe monomer, 2-body and 3-body interactions with
a classical empirical polarization term for higher order many-body
effects.\\

\noindent
The goal of the current work is to combine accurate neural network
(NN) intramolecular energies from PhysNet\cite{MM.physnet:2019} with
molecular mechanics ``(k)MDCM"\cite{MM.mdcm:2017,MM.kmdcm:2024}
zeroth-order electrostatics and a new 2-body potential to handle
non-bonded interactions at short-range. PhysNet intramolecular
contributions offer a reliable description of the monomer PES, while
(k)MDCM + Lennard-Jones (LJ) terms provide similar accuracy for medium
to long-range non-bonded interactions at reduced computational
cost. The remaining, short-range intermolecular interactions are
described by adding PhysNet dimer potentials, maintaining a similar
level of accuracy to the intramolecular and long-range terms, but
within a 2-body approximation.\cite{MM.ffs:2024} Missing many-body
effects will be added in future work. \\

\noindent
The combination of methods envisaged capitalizes on the strengths of
each component part. A simple hybrid ML/MM approach that combined ML
monomer potentials with classical force field non-bonded terms at all
intermolecular separations will suffer from reduced accuracy in
condensed-phase simulations due to the coarse approximation of force
field non-bonded terms at short-range.\cite{MM.ffs:2024} Similarly,
training pure ML models for high-dimensional, heterogeneous systems
leads to difficulties in generating sufficient training data and
requires larger neural networks for accurate description of that data,
increasing computational cost in evaluating energies and forces in
molecular dynamics (MD) simulations. Explicit dimer potentials offer a
compromise in the short-range, non-bonded region where they are more
accurate than force fields within the limits of the 2-body
approximation,\cite{MM.ffs:2024} and they require substantially less
overhead to train than high-dimensional ML potentials.\\

\noindent
The preference for a 2-body ML potential over an empirical force field
with improved non-bonded terms such as penetration
corrections\cite{stone:2011} or
charge-transfer\cite{misquitta:2013,gresh:1986} arises from the nature
of non-bonded interactions at close range. Monomer wave function
overlap cannot be neglected at shorter intermolecular separations,
leading to reorganization of electron clouds in accordance with Pauli
exclusion in ways that are difficult to describe using simple
empirical potentials that do not account for electrons explicitly. In
addition, the available conformational and permutational space of
non-bonded neighbors is much larger than for bonded interactions where
neighbors are fixed, meaning such empirical potentials must also be
particularly transferable, leading to much work fine-tuning the
functional
forms.\cite{stone:2011,Naseem-Khan:2022,ren:2003,Wu:2019}.\\

\noindent
Long-range interactions, on the other hand, are dominated by
electrostatic Coulomb terms as overlap of the monomer wave functions
is small.\cite{murrell:1965} These interactions are well described
using classical electrostatics models such as assigned point charges
or multipole expansions of the monomer charge
densities,\cite{MM.mdcm:2017,stone:1985,ponder:2010,cardamone:2014}
coupled with higher order terms such as
polarization\cite{ren:2003,Thole1981,MM.mdcm:2020,lemkul:2016,Lagardere:2017}
and dispersion corrections\cite{Naseem-Khan:2022} as needed.\\

\noindent
As a consequence, fitting explicit ML dimer surfaces for each
interacting pair avoids problems by precalculating these energy
surfaces explicitly for each interaction type, increasing
parametrization costs with respect to an empirical force field but
yielding accurate pairwise energies for all conformations and
interaction permutations. \\

\noindent
The solution pursued here is therefore to apply NNs at short range to
overcome the difficulties and computational costs associated with
empirical interaction potentials in this region, switching to more
physics-inspired electrostatic interactions combined with LJ-models
beyond a cutoff at longer range where these models perform well and at
significantly reduced computational cost. Initially, this approach was
applied to dichloromethane (DCM), for its challenging electrostatics
involving H-bonding and sigma holes,\cite{Almasy:2019,allen:2013} but
also as DCM is an ideal test system for a 2-body potential where
many-body effects have been found to play a minor
role.\cite{MM.ffs:2024} These results are then contrasted with
acetone, which provides a counter-example where missing many-body
contributions would need to be included, as a motivation for future
work. Finally, exploratory molecular dynamics simulations are carried
out to establish the utility and feasibility of the total interaction
model for dynamics studies. \\

\section{Methods}

\subsection{Data Generation}
All condensed-phase simulations to generate reference structures were
carried out using the CHARMM suite of codes.\cite{Charmm-Brooks-2009}
The simulation systems consisted of a $32^3$ \AA\/$^3$ cubic box
generated using PACKMOL.\cite{martinez.pm:2009} For DCM and acetone
the systems contained 320 and 300 monomers, respectively. Both systems
were initially relaxed using 2000 steps of Steepest Descent (SD)
minimization. This was followed by heating and equilibration
simulations of 20 ps and 50 ps, respectively, with a timestep of
$\Delta t = 1$ fs, in the $NVT$ and $NpT$ ensembles applying
SHAKE\cite{ryckaert.1977} on bonds involving H-atoms. A Nose-Hoover
thermostat\cite{Evans:1985,nose:1984,nose:2002,hoover:1985} was used
to maintain the temperature at 300 K. Finally, 1 ns production
simulations in the $NpT$ ensemble were carried out using the Leapfrog
Verlet integrator with a timestep of $\Delta t = 0.2$ fs whereby all
bonds involving hydrogen atoms were now free to move. The pressure was
controlled using the Langevin Piston barostat\cite{Feller:1995} with a
reference pressure of 1 atm. Long range interactions were treated
using particle mesh Ewald\cite{feller:1996} with a cutoff of 14 \AA\/
and the Lennard-Jones interactions were switched between 10 \AA\/ and
12 \AA\/. A total of 200 and 100 snapshots were saved for DCM and
acetone, respectively.\\

\subsection{The MDCM Models}
For the conformationally averaged minimal distributed charge models
(MDCMs)\cite{MM.mdcm:2017,MM.mdcm:2020,MM.kmdcm:2024} first the
structure of each monomer was geometry-optimized at the
MP2/aug-cc-pVTZ level of theory using Gaussian16.\cite{gaussian16}
Following this, additional structures were generated from MD
simulations using ASE\cite{larsen:2017} and the XTB semiempirical
functional\cite{bannwarth.XTB:2019} and 5 structures were selected
from a 1 ns run using a timestep of 0.1 fs, from a set of 10000
structures.\\

\noindent
Next, the molecular electrostatic potential (ESP) was computed across
a grid for all 6 structures at the MP2/aug-cc-pVTZ level of theory. A
grid resolution of $M_x \times M_y \times M_z$ reference points
$(x_k,y_l,z_m)$ with $\{ k,l,m \} = 1 \dots \{ M_x,M_y,M_z \}$ was
used, on which the electron density and ESP of the converged
wavefunction was determined. Points within the molecular volume, as
defined by the van der Waals radii of each atom, were discarded. The
grid spacing on which the ESP was evaluated was 0.33 a$_0$ and 0.19
a$_0$ for DCM and acetone, respectively. Hence, the reference data
consists of discrete values for ESP$_{\rm ref}(x,y,z)$ on a regular
grid. This target quantity was then represented by a minimal
distributed charge model (MDCM) with $N$ off-center point charges
$\{q_k,\mathbf R_k\}_{k=1}^N$. The charge magnitudes $q_k$ and their
position ${\rm \bf R}_k$ relative to the reference atom are fitted
such as to best reproduce the reference ESP$(x,y,z)$ by minimizing the
loss function $\mathcal{L}_j = \sqrt{ \frac{1}{N_{\rm pts}} \sum_{ i =
    1}^{N_{\rm pts}} [{\rm ESP}_{{\rm ref},j} ({\bf r}_{i}) - {\rm
      ESP}_{{\rm MDCM},j}({\bf r}_{i})]^2}$ subject to $\sum_k
q_k=Q_{\mathrm{tot}}$, where ${\bf r}_{i}$ is the position vector of
each of the $N_{\rm pts}$ grid points outside the molecular
volume. For a conformationally averaged MDCM model comprising $N_{\rm
  conf}$ conformers, $\mathcal{L} = \sum_{j=1}^{N_{\rm conf}}
\mathcal{L}_j$ was the loss function. Minimization of the loss
function was carried out using differential
evolution\cite{Storn.DE:1997,MM.dcm:2014,MM.mdcm:2017,MM.mdcm:2020}.\\

\noindent
To determine the minimum number of charges $N$ required, models were
generated for each $N \in [N_{\rm min},N_{\rm max}]$ charges,
recording the lowest $\mathrm{RMSE}(N)$ from several fits at each
$N$. When the improvement in the RMSE achieved by increasing $N$
dropped below $\sim 0.1$ kcal/mol, fitting was halted and the model
was selected. Otherwise, the value of $N_{\rm max}$ was increased
further and the DE fit was repeated. In the present work, 14 and 16
charges were sufficient for DCM and acetone, respectively.\\

\subsection{The ML-Based Energy Functions}
The reference data for training the NN-PESs were determined using
ORCA\cite{ORCA,ORCA5} with DLPNO-MP2\cite{pinski:2015,neugebauer:2023}
and cc-pVTZ and cc-pVDZ basis sets basis for DCM and for acetone,
respectively. For DCM and acetone 4000 and 2000 monomer geometries
were extracted from the 200 and 100 distinct clusters,
respectively. Furthermore, (DCM)$_2$ and (acetone)$_2$ dimer
structures were extracted from the 200 and 100 distinct clusters,
respectively. Each 20mer cluster gives rise to $\binom{20}{2} =
\frac{20!}{2!(20-2)!}  = \frac{20 \cdot 19}{2 \cdot 1} = 190$
different dimers. In total, this yields 38000 (DCM)$_2$ and 19000
(acetone)$_2$ dimer structures. In addition to the DLPNO-MP2 monomer
and dimer reference data (energies and forces), total energies for the
20mer clusters were also determined. This information was used to (i)
assess the accuracy of total NN-PES cluster energies and cluster
energies from empirical energy functions and (ii) quantify the
residual many-body contributions beyond pairwise interactions. No BSSE
corrections were included.\\

\noindent
PhysNet models \cite{MM.physnet:2019} were trained jointly on monomers
and dimers for each species (DCM and acetone) using the Asparagus
environment.\cite{mm:asparagus2025} For DCM the dataset consisted of
38000 dimers and 4000 monomers, whereas the acetone dataset consisted
of 19000 dimers and 2000 monomers. For training, both datasets were
split into 80/10/10 for the training, validation and test sets,
respectively. The models were trained on $E^{\text{ref}}$ (total
energies), $F^{\text{ref}}_{i,\alpha}$ (forces), $Q^{\text{ref}}$
(charges), and $p^{\text{ref}}_{\alpha}$ (dipole-moment components)
with reference calculation done at the DLPNO–MP2 level in ORCA
\cite{ORCA,ORCA5}, using cc-pVTZ for DCM and cc-pVDZ for acetone. The
loss function to be optimized was
\begin{equation}
  \label{eq:loss1}
\begin{aligned}
\mathcal{L} &= w_E \left| E - E^{\text{ref}} \right| +
\frac{w_F}{3N_{\rm a}} \sum_{i=1}^{N_{\rm a}} \sum_{\alpha=1}^{3}
\left| -\frac{\partial E}{\partial r_{i,\alpha}} -
F^{\text{ref}}_{i,\alpha} \right| \\ &+ w_Q \left| \sum_{i=1}^{N_{\rm
    a}} q_i - Q^{\text{ref}} \right| + \frac{w_p}{3}
\sum_{\alpha=1}^{3} \left| \sum_{i=1}^{N_{\rm a}} q_i r_{i,\alpha} -
p^{\text{ref}}_{\alpha} \right| + \mathcal{L}_{\text{nh}}.
\end{aligned}
\end{equation}
where $N_{\rm a}$ is the number of atoms in the species (DCM or
acetone) considered. The Adam optimizer\cite{kingma:2014,reddi:2019}
was used for minimizing $\mathcal{L}$ and the
hyperparameters\cite{MM.physnet:2019,MM.physnet:2023} $w_i$ for $i \in
{E,F,Q,p}$ weight the contributions to the loss and were $w_E=1$,
$w_F\approx 52.92$, $w_Q\approx 14.39$, and $w_p\approx 27.21$.  The
term $\mathcal{L}_{\text{nh}}$ is the “non-hierarchical penalty” used
as a regularizer \cite{MM.physnet:2019}. The resulting fitted charges
and dipole moments were not used here.\\

\subsection{Adjusting the Lennard-Jones Parameters}
The nonbonded interaction in an empirical energy function comprises
Lennard-Jones and electrostatic contributions for which either a point
charge (PC) or an MDCM description was used in the present work. Two
different strategies were pursued to improve the empirical FFs by
adjusting the Lennard-Jones (LJ) parameters. ``Approach A" assumes
that the total nonbonded interaction energy of a cluster consists of
pairwise additive dimer interactions which necessarily neglects
many-body contributions beyond the dimer, whereas for ``Approach B"
the total cluster nonbonded interaction was employed as the target
quantity which does contain many-body contributions.\\

\noindent
{\it Approach A:} The total energy $E^{\mathrm{pair}}_j$ for cluster
$j$ is the sum of all pairwise contributions (dimers)
\[
E^{\mathrm{pair}}_j = \sum_{i=1}^{190} E^{\mathrm{dimer,inter}}_{i,j}.
\]
For optimizing the LJ-parameters, the force-field non-bonded
interaction energies for all clusters $j$, $E_{{\rm FF},j}^{\rm
  pair}$, were compared with the sum of reference dimer energies,
$E_{{\rm ref},j}^{\rm pair}$ at the DLPNO--MP2/[cc-pVTZ / cc-pVDZ]
level of theory for each of the 190 dimers comprising cluster $j$ for
DCM and acetone, respectively.\\

\noindent
{\it Approach B:} In this case, the total energy for each cluster from
the empirical energy function, $E_{{\rm FF},j}^{\rm full}$, and
$E_{{\rm ref},j}^{\rm full}$ from reference DLPNO--MP2/[cc-pVTZ /
  cc-pVDZ] calculations was employed for adjusting the
LJ-parameters. Note that $E_{{\rm ref},j}^{\rm full}$ contains
many-body contributions beyond the dimer whereas $E_{{\rm ref},j}^{\rm
  pair}$ from ``Approach A" does not.\\

\noindent
In both approaches, the {\it ab initio} reference energies $E_{{\rm
    ref},j}^{\rm pair}$ and $E_{{\rm ref},j}^{\rm full}$,
respectively, were used as the target for fitting the total nonbonded
interaction energies of each cluster $j$ represented as an empirical
energy expression. Thus, the Lennard-Jones parameters $\epsilon$ and
$R_{\rm min}/2$ for each atom type were optimized. Following the
conventions in the CHARMM and CGenFF force fields,
Lorentz-Berthelot\cite{berthelot:1898,lorentz:1881} combination rules
were used for combining LJ-parameters of different atom types. During
fitting, each LJ-parameter was constrained towards the original CGenFF
values. The force field interaction energy $E_{{\rm FF},j} = E^{\rm
  elec}_{j} + E^{\rm LJ}_{j}$ was the sum of electrostatic and LJ
contributions and the loss function to be minimized was the difference
between $E_{{\rm FF},j}$ and $E_{{\rm ref},j}$
\begin{equation}
\mathcal{L}
= \frac{1}{\mathcal{J}}\sum_{j\in\mathcal{J}}
\Big( E_{{\rm FF},j} - E_{{\rm ref},j} \Big)^2
\label{eq:loss2}
\end{equation}
For approaches A and B the interaction energies were either sums of
dimer energies or total cluster energies as explained above,
$\mathcal{J}$ is the total number of clusters and the optimisation was
performed using a Nelder-Mead\cite{nelder:1965} algorithm as
implemented in the \emph{Scipy\cite{virtanen:2020}} module in
Python.\\

\subsection{The ML/MM Boundary}
\begin{figure}[h!]
\centering \includegraphics[width=0.75\textwidth,
  height=0.42\textheight]{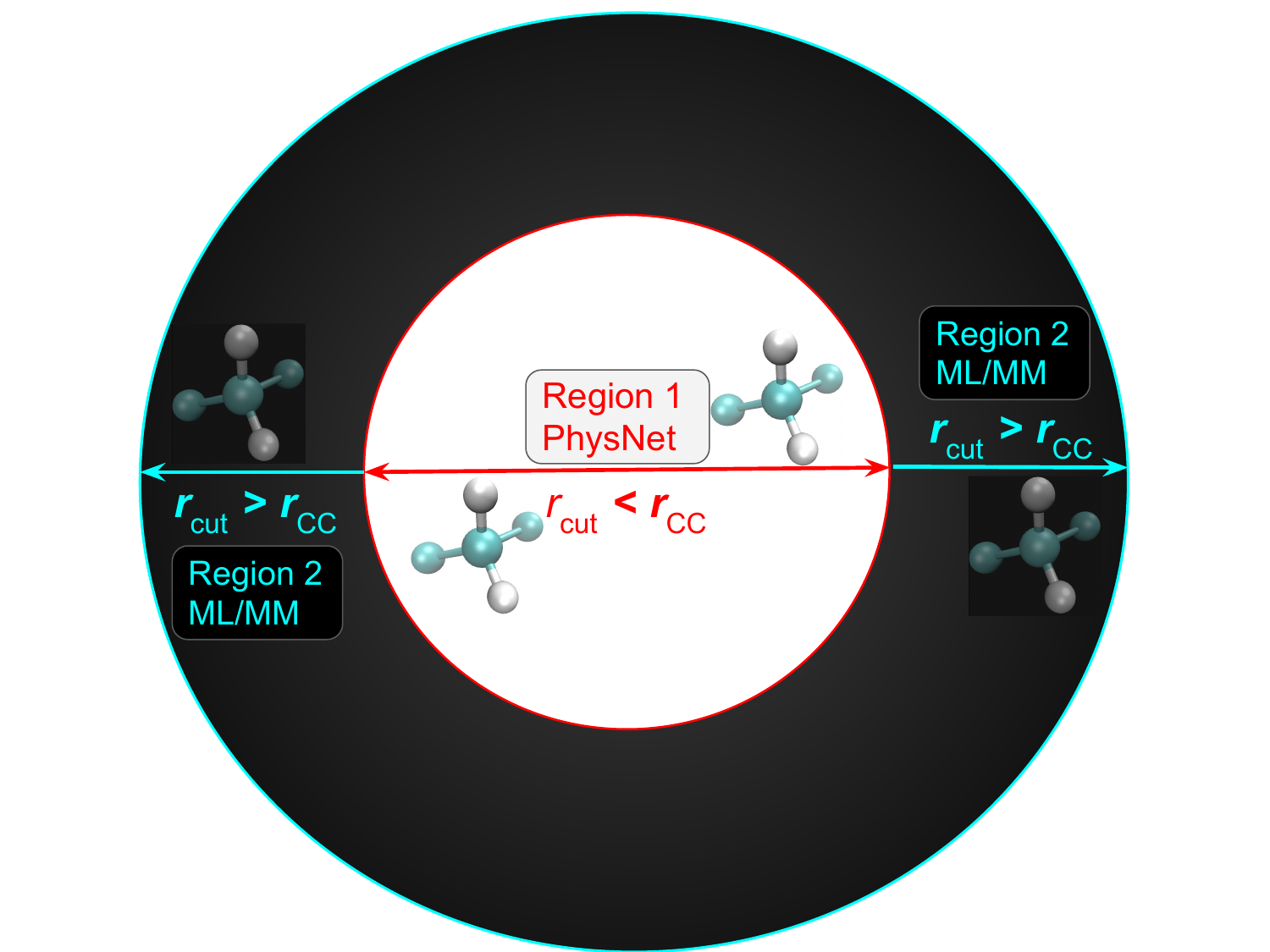}
\caption{Definitions of ``Region 1" and ``Region 2" necessary to model
  the ML/MM boundary. }
\label{fig:fig1} 
\end{figure}

A hybrid ML/MM energy function with explicit two-body ML-interactions
requires at least two regions. Interactions with monomers within
``Region 1" around a first monomer are evaluated using the ML-PES. For
interactions with monomers within ``Region 2" an MM-description is
employed, see Figure \ref{fig:fig1}. In the present work, the
MM-energy function uses either charges from the CGenFF force field or
MDCM-charges combined with CGenFF LJ-parameters or those refitted to
best describe the quantum chemical reference data.\\

\noindent
For the hybrid ML/MM energy function the trained PhysNet model was
used for $r_i < r_{\rm cut}$, whereas for $r_i \ge r_{\rm cut},$ the
energy was computed from the MM energy function. In the following,
$r_i$ is the C--C distance for DCM and the carbonyl C--C distance for
acetone. For a given value of $r_{\rm cut}$, the total nonbonded
interaction for a collection of monomers is $E_{\rm MM}$ for $r_i \geq
r_{\rm cut}$ and $E_{\rm ML} = E_{\rm ML}^{\rm dimer} - E_{\rm
  ML}^{\rm monomer A} - E_{\rm ML}^{\rm monomer B}$ for $r_i < r_{\rm
  cut}$. Sensitivity of this hybrid ML/MM energy function with respect
to the actual value of $r_{\rm cut}$ was probed for values $3 \leq
r_{\rm cut} \leq 18$~\AA\/ in steps of 1~\AA\/.\\

\subsection{Molecular Dynamics Simulations}
Finally, the applicability of the potential energy functions developed
here to molecular dynamics simulations was demonstrated.  For this,
gas phase clusters of increasing sizes ($n \in \{ 2, 10, 20\}$) were
generated using PACKMOL, where different initial (local) densities
were obtained using different box side lengths for the initial
structure generation ($\L \in \{ 10.0, 20.0, 30.0, 40.0\}$). Cluster
simulations were carried out in the $NVE$ ensemble using ASE,
pyCHARMM, and jaxmd.\cite{JAXMD2020} Integration time steps (ranging
from 0.05 to 0.5 fs) were tested to check changes in the variance of
the total energy. Similarly, the variance $(E - <E>)^2$ of the total
energy for simulations $\sim 1$ ns in length was followed and the
influence of training and using single- vs. double-precision
(\textit{float32} (the \textit{jax} default) to \textit{float64}) for
running the simulations was assessed. This was motivated by recent
findings that depending on the observable sought, increased numerical
precision is mandatory for meaningful results and may lead to more
stable simulations.\cite{MM.acc:2024}\\

\noindent
For MD simulations, the cut-offs for switching between the ML and MM
regions were implemented based on the SO3LR
potential.\cite{kabylda:2025} Three parameters are available: the
position of the switching region (ML to MM) and the width $\sigma_{\rm
  ML}^{\rm width}$ of the ML switching regime (smaller is better) and
half the width of the MM region, where the potential is damped
(smoothly and continuously) such that the derivative at the bounds of
the cut-off is equal to 0.0 (kcal/mol)/\AA\/.  The contribution of the
ML potential energy surface (ML-PES) is controlled by a smooth
switching function $v_{\rm switch}(r)$ defined in terms of an
auxiliary variable $s(r)$ that varies between 0 and 1.  The function
$s(r)$ was defined as
\begin{equation}
s(r) =
\begin{cases}
0, & r \le r_0, \\[6pt]
\dfrac{r-r_0}{r_1-r_0}, & r_0 < r < r_1, \\[10pt]
1, & r \ge r_1,
\end{cases}
\end{equation}
where $r_0$ and $r_1$ define the inner and outer boundaries of the
switching region. A smooth $C^1$ switching profile is then constructed
as
\begin{equation}
v_{\rm switch}(r) = s(r)^2\left(3 - 2s(r)\right).
\end{equation}

\noindent
The total interaction energy is obtained by smoothly interpolating
between the ML and MM descriptions,
\begin{equation}
V(r) = v_{\rm switch}(r) V_{\rm ML}(r) +
\left[1 - v_{\rm switch}(r)\right] V_{\rm MM}(r).
\end{equation}
For the MM region, interactions are gradually turned off using a
distance-dependent cutoff applied to the center-of-mass separation
$r_{\rm CoM}$ between monomers. The cutoff is defined as
\begin{equation}
v_{\rm MMcut} = v_{\rm switch}\!\left( \frac{r_{\rm CoM} - [\rho_{\rm
      MM}^{\rm on} + \sigma_{\rm ML}^{\rm width}]} {\rho_{\rm MM}^{\rm
    on} - \sigma_{\rm ML}^{\rm width}} \right)^{\gamma},
\end{equation}
where $\rho_{\rm MM}^{\rm on}$ defines the distance at which the MM
interaction begins to be switched off, $\sigma_{\rm ML}^{\rm width}$
controls the width of the transition region, and $\gamma = 1$
determines the steepness of the cutoff see Figure \ref{sifig:fig7}.\\

\noindent
It is noted that the C--C separation used as the progression
coordinate for defining the boundary for the ML/MM potential slightly
differs from $r_{\rm CoM}$ used for the MD simulations. This was done
for practical reasons when fitting the LJ-parameters and is no
limitation. Using $r_{\rm CoM}$ for the MD simulations is considered
more general.\\

\section{Results}
The following section describes validation of the machine learned
PhysNet dimer PESs (ML-PESs), followed by assessing the performance of
CGenFF and the fitted CGenFF+MDCM models after refitting LJ parameters
to the reference data. The accuracy of the different non-bonded
contributions as a function of intermolecular separation is then used
to select an appropriate cut-off to switch between the ML-PES, which
performs well at all ranges, and CGenFF or CGenFF+MDCM, which perform
well at medium to long range.  Finally, the stability of an initial
implementation of this 2-body potential to perform MD simulations is
shown for $NVE$ simulations on the ns timescale. \\

\subsection{Validation of Monomer and Dimer ML-PESs}
\begin{figure}[H]
\centering \includegraphics[width=1.005\textwidth,
  height=0.4\textheight]{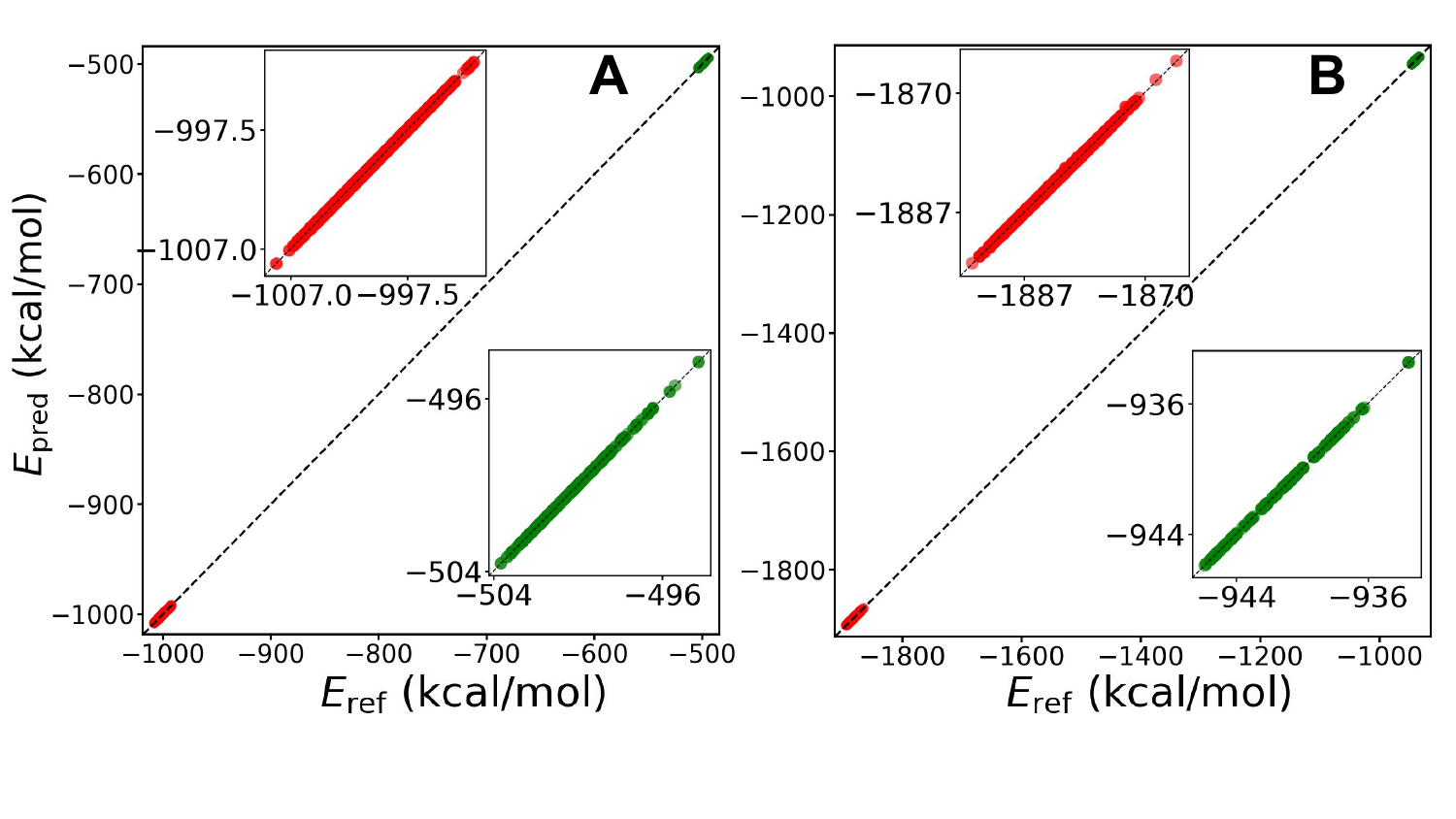}
\caption{Training performance of Dichloromethane and Acetone evaluated
  on the test set. Panel A: Performance of the PhysNet model for
  DCM. Panel B: Performance of the PhysNet model for acetone. The
  green and red symbols are for monomer and dimer energies,
  respectively. For both models the correlation coefficient is $R^{2}
  = 0.99$.}
\label{fig:fig2} 
\end{figure}

To set the stage, the machine-learned monomer and dimer PESs were
validated. This was done by determining the performance of the trained
PhysNet models on the test set which contained both, monomer and dimer
structures. Figure \ref{fig:fig2} demonstrates that the monomer and
dimer contributions for both systems are very accurately represented
compared with the reference DLPNO-MP2/cc-pVTZ and DLPNO-MP2/cc-pVDZ
data for DCM and acetone, respectively. For both systems monomer
deformation and dimer interaction energies cover a range of 10
kcal/mol.\\

\noindent
For DCM PhysNet was trained on 42000 structures (4000
monomer and 38000 dimer geometries) comprising $(E_{i}, F_{i}, \mu_{i}
= q r_{i}, q_{i})$ using Asparagus\cite{mm:asparagus2025}. For DCM the
test set RMSE-values for monomers and dimers on the jointly fitted
model are 0.0114 kcal/mol and 0.0282 kcal/mol, respectively, whereas
the total RMSE is 0.0269 kcal/mol. This compares with 0.0161 kcal/mol
and 0.0325 kcal/mol for individually trained monomer and dimer
models. For acetone these values are 0.0119 kcal/mol and 0.0628
kcal/mol for monomer and dimer for the jointly fitted model and 0.0599
kcal/mol for the total RMSE. The individually trained monomer and
dimer models feature RMSE values of 0.0509 kcal/mol and 0.0585
kcal/mol, respectively.\\

\noindent
The performance of all trained models discussed above is within the
expectations of previous work. For example, trained PhysNet models for
{\it syn-}Criegee based on MP2 calculations reported RMSE$(E) = 0.19$
kcal/mol for a reactive potential energy surface.\cite{upadhyay:2021}
Similarly, PhysNet trained on MP2-reference data for malonaldehyde and
singly and doubly methylated Malonaldehyde reported RMSE-values of
0.0005 to 0.0218 kcal/mol depending on the substitutions
considered.\cite{MM.tl:2021} Comparable RMSE$(E)$-values ranging from
0.020 to 0.042 kcal/mol were reported for long alkane chains trained
using the t-MB-PIP framework\cite{Richardson.2025}.\\

\subsection{Two-Body, Range-Separated ML/MM}
As mentioned in the methods, two approaches were chosen to readjust
the CGenFF LJ-parameters to the reference data and to using MDCM
instead of PC-electrostatics. This sections discusses the performance
of ``Approach A" which employed dimer interaction energies as the
reference to fit to. The performance of ``Approach B", which uses
total cluster interaction energies, is discussed in the final
Section.\\

\begin{figure}[h!]
\centering \includegraphics[width=0.8\textwidth,
  height=0.60\textheight]{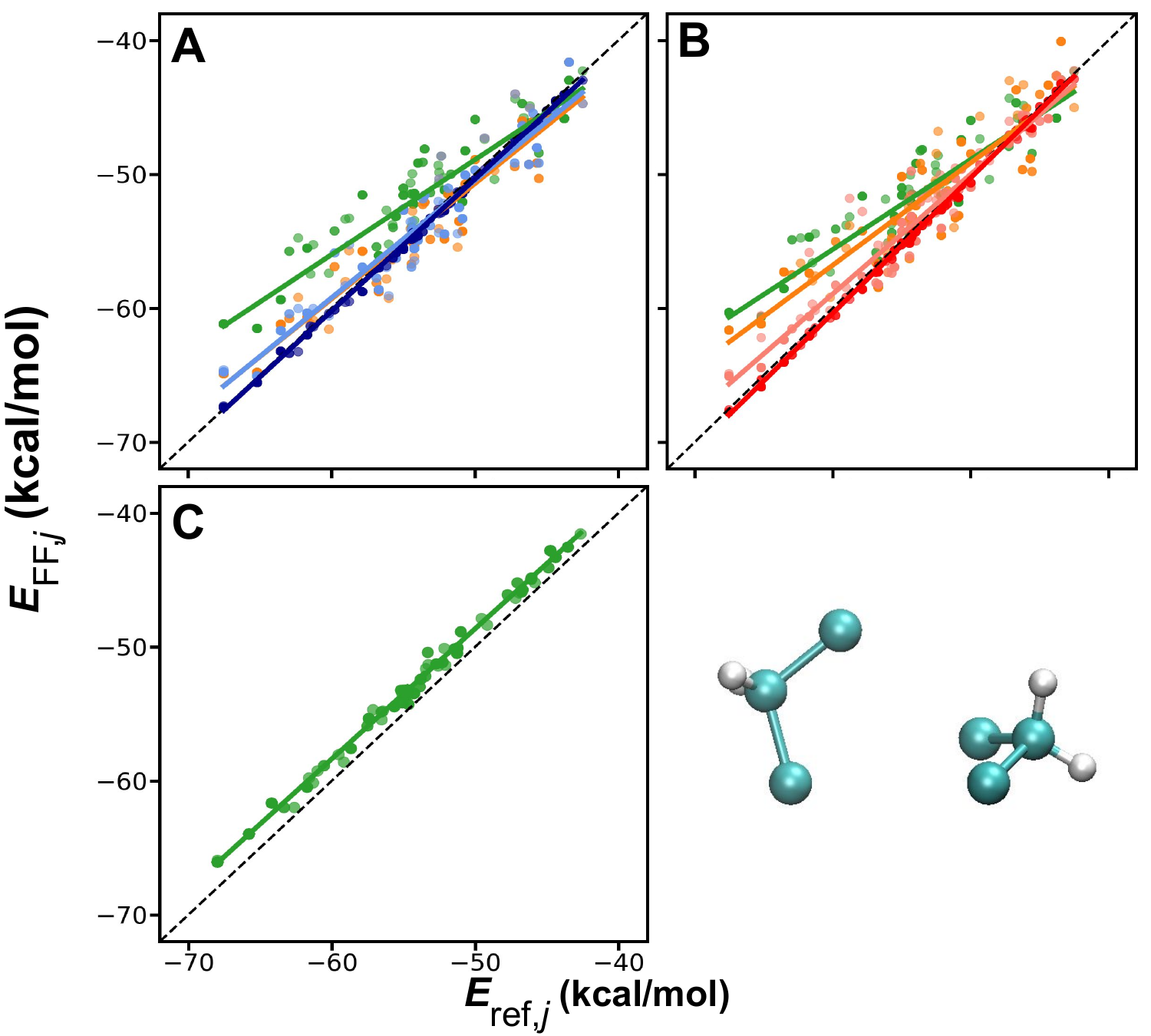}
\caption{Correlation between reference electronic structure data
  ($x-$axis) and the MM (panels A and B) or ML (panel C) energies for
  DCM using ``Approach A''. In panels A to C, the green traces are
  results from using CGenFF, CGenFF+MDCM, and PhysNet energies without
  modification. The orange traces in panels A and B are the results
  after refitting the LJ-parameters but without ML-contributions. The
  performance of hybrid ML/MM models are the blue and red symbols and
  lines in panels A and B, respectively. Light and dark traces are for
  $r_{\rm cut} = 4$ \AA\/ and $r_{\rm cut} = 7$ \AA\/,
  respectively. Note that for larger values of $r_{\rm cut}$, the
  mixed ML/MM models progressively approach the PhysNet
  model. Computationally more advantageous are ML/MM models with
  shorter $r_{\rm cut}$. For statistical performance measures, see
  Table \ref{tab:app-a}.}
\label{fig:fig3} 
\end{figure}

\begin{figure}[h!]
\centering
\includegraphics[width=0.8\textwidth]{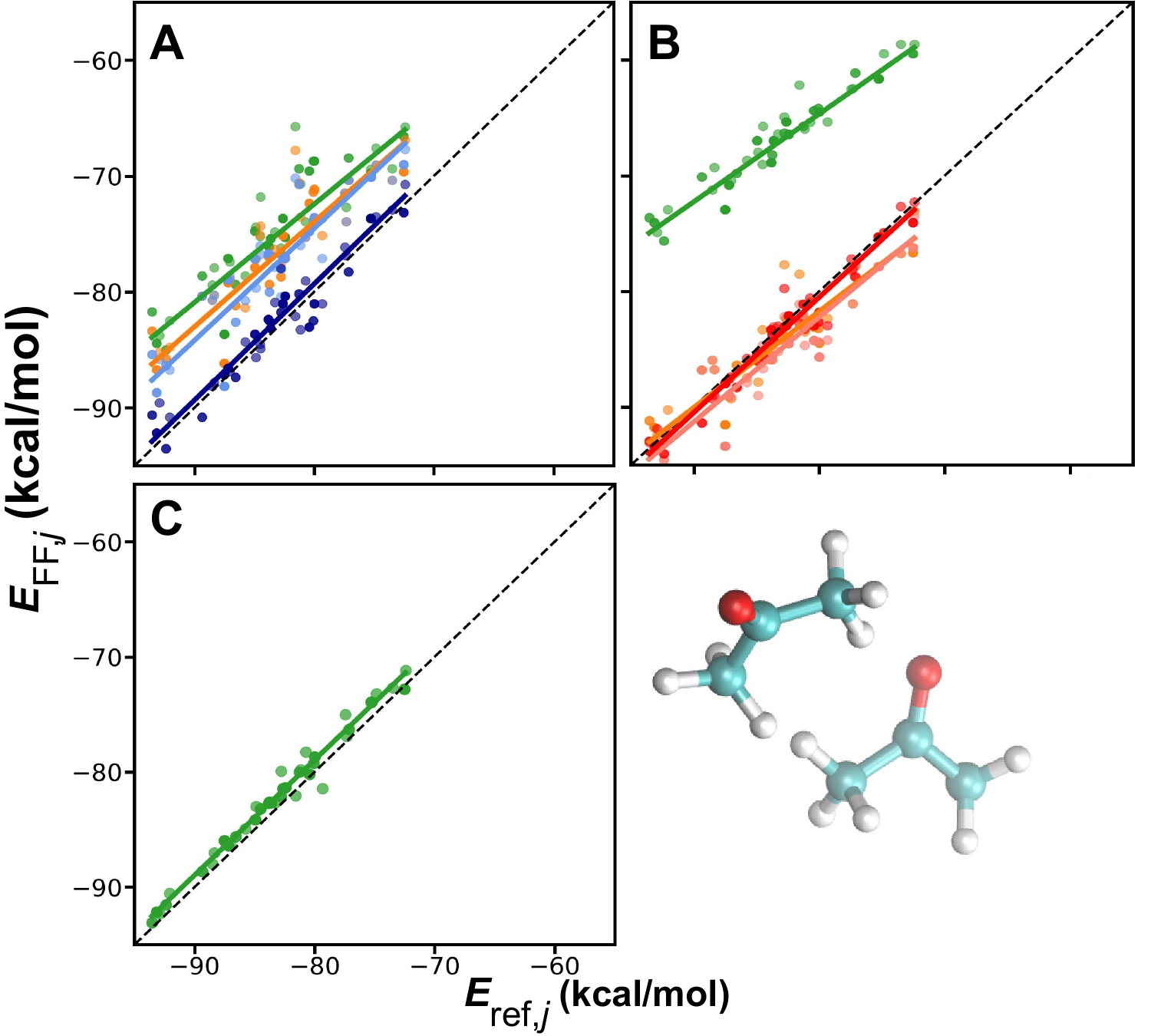}  
\caption{Correlation between reference electronic structure data
  ($x-$axis) and the MM (panels A and B) or ML (panel C) energies for
  acetone using ``Approach A''. In panels A to C, the green traces are
  results from using CGenFF, CGenFF+MDCM, and PhysNet energies without
  modification. The orange traces in panels A and B are the results
  after refitting the LJ-parameters but without ML-contributions. The
  performance of hybrid ML/MM models are the blue and red symbols and
  lines in panels A and B, respectively. Light and dark traces are for
  $r_{\rm cut} = 4$ \AA\/ and $r_{\rm cut} = 7$ \AA\/,
  respectively. Note that for larger values of $r_{\rm cut}$, the
  mixed ML/MM models progressively approach the PhysNet
  model. Computationally more advantageous are ML/MM models with
  shorter $r_{\rm cut}$. For statistical performance measures, see
  Table \ref{tab:app-a}.}
\label{fig:fig4} 
\end{figure}

\noindent
Using ``Approach A", the 2-body cluster energies are the reference
data to obtain refined LJ-parameters for CGenFF and CGenFF+MDCM, see
Figures \ref{fig:fig3} and \ref{fig:fig4} and Tables \ref{sitab:tab3}
and \ref{sitab:tab4}. Without readjusting the LJ-parameters (green
symbols and lines), the [RMSE$(E)$, MAE$(E)$] are [3.45, 2.83]
kcal/mol and [8.30, 9.93] kcal/mol for DCM and acetone,
respectively. Because CGenFF is based on MP2/6-31G(d) calculations for
dimers and fitting to condensed phase properties from simulations for
the LJ-parameters,\cite{cgenff:2012} allowing the LJ ranges and well
depths to vary for the DLPNO-MP2/cc-pVTZ ad DLPNO-MP2/cc-pVDZ
reference data is still expected to improve the model. Evidently, when
changing from a point charge-based description to MDCM, the
LJ-parameters also need to be optimized.\\

\noindent
As can be seen in Figures \ref{fig:fig3} and \ref{fig:fig4} (orange
symbols and lines) and quantified in Table \ref{tab:app-a} for DCM and
acetone, the refitted parameters (Table \ref{tab:lj}) primarily lead
to a systematic shift in the CGenFF and CGenFF+MDCM cluster energies,
moving the mean of the distributions closer to the mean of the
reference DLPNO-MP2 2-body energies. This reduces the RMSE from 3.45
to 1.88 kcal/mol (CGenFF) and from 3.74 to 3.26 kcal/mol (CGenFF+MDCM)
in the case of DCM. Relaxing constraints on allowed LJ parameter
values would further remove the residual systematic shift for
(CGenFF+MDCM), but typically without improving the residual slope and
variance. For acetone, [RMSE$(E)$, MAE$(E)$] reduce from [8.30, 7.93]
to [6.68, 6.27] kcal/mol for CGenFF and from [16.01, 15.92] to [2.38,
  2.07] kcal/mol for CGenFF+MDCM, respectively. Again, residual shift
for CGenFF could be removed by further relaxing constraints on
parameter values, but likely without improving the residual variance
and with a danger of overfitting at the expense of accuracy in regions
of conformational space that are not well represented in the training
data. The consistent observation that residual variance (the spread in
the data away from the diagonal) is not substantially reduced by
refitting highlights the errors that remain even after refitting LJ
terms if such force fields are used for all non-bonded interactions in
condensed phase. In contrast, the PhysNet ML-PES (green symbols and
lines in panel C) achieves much lower variance, faithfully
representing the 2-body reference data with an RMSE of 1.07 kcal/mol
and 1.20 kcal/mol for DCM and acetone, respectively. The small
residual shift from PhysNet results from cumulative errors when
summing over all 190 ML dimer pairs in each cluster, while the LJ
parameters for MM models are deliberately fitted to summed pairwise
cluster energies to avoid cumulative errors.\\

\begin{table}[h!]
\centering
\small
\renewcommand{\arraystretch}{1.2}
\begin{tabular}{lcccc|cccc}
\toprule
& \multicolumn{4}{c|}{\textbf{Dichloromethane (DCM)}} & \multicolumn{4}{c}{\textbf{Acetone}} \\
\cmidrule(lr){2-5}\cmidrule(lr){6-9}
\textbf{Model} & $R^2$ & RMSE & MAE & STD & $R^2$ & RMSE & MAE & STD \\
\midrule
CGenFF              & 0.92 & 3.45 & 2.83 & 2.55 & 0.81 & 8.30 & 7.93 & 2.45 \\
CGenFF + Fit        & 0.93 & 1.88 & 1.55 & 1.87 & 0.83 & 6.68 & 6.27 & 2.29 \\
ML/MM (4 \AA)       & 0.95 & 1.70 & 1.40 & 1.70 & 0.84 & 6.04 & 5.64 & 2.30 \\
ML/MM (7 \AA)       & 0.99 & 0.40 & 0.35 & 0.26 & 0.94 & 1.53 & 1.32 & 1.37 \\
\cmidrule(lr){1-9}
CGenFF+MDCM         & 0.90 & 3.74 & 3.00 & 2.78 & 0.93 & 16.01 & 15.92 & 1.77 \\
CGenFF+MDCM + Fit   & 0.87 & 3.26 & 2.69 & 2.69 & 0.88 & 2.38 & 2.07 & 1.98 \\
ML/MM (4 \AA)       & 0.96 & 1.52 & 1.22 & 1.49 & 0.88 & 2.79 & 2.30 & 1.96 \\
ML/MM (7 \AA)       & 0.99 & 0.42 & 0.34 & 0.27 & 0.97 & 1.08 & 0.73 & 0.96 \\
\cmidrule(lr){1-9}
PhysNet             & 0.99 & 1.07 & 1.01 & 0.50 & 0.98 & 1.20 & 1.11 & 0.59 \\
\bottomrule
\end{tabular}
\caption{Error statistics for DCM and acetone following ``Approach A"
  for optimizing the LJ-parameters, see orange symbols and lines in
  Figures \ref{fig:fig3}A/B and \ref{fig:fig4}A/B. Values for $R^2$,
  RMSE, MAE, and STD (in kcal/mol) for correlations between reference
  energies $\sum_{i=1}^{190} E^{\rm dimer}_{i}(\rm Ref)$ and energies
  using unfitted and fitted LJ-parameters $\sum_{i=1}^{190} E^{\rm
    dimer}_{i}(\rm Model)$. Results for PhysNet and the corresponding
  ML/MM models using cutoff values of $r_{\rm cut}=4$ and $7$~\AA \/
  are also reported for comparison.}
\label{tab:app-a}
\end{table}

\begin{table}[h!]
\centering
\caption{Original and fitted Lennard--Jones parameters for
  Dichloromethane and Acetone atom types using CGenFF and CGenFF+MDCM
  approaches for DCM and acetone. For both, ``Approach A" and
  ``Approach B" no cutoff was used.}  {\scriptsize
\begin{tabular}{l|l|lc||cc|cc}
\hline
Molecule & Atom Type & Parameter & CGenFF &
\multicolumn{2}{c|}{CGenFF+Fit} &
\multicolumn{2}{c}{CGenFF+MDCM+Fit} \\
 &  &  &  &
Approach A & Approach B &
Approach A & Approach B \\
\hline
\multirow{6}{*}{Dichloromethane}
& CG321 (C) & $\epsilon$     & -0.0560  & -0.0839 & -0.0627 & -0.0840 & -0.0784 \\
&           & $R_{\min}/2$   &  2.0100  &  1.8478 &  1.8590 &  1.8180 &  1.8957 \\
& HGA2 (H)  & $\epsilon$     & -0.0350  & -0.0524 & -0.0420 & -0.0525 & -0.0490  \\
&           & $R_{\min}/2$   &  1.3400  &  1.1956 &  1.2559 &  1.2314 &  1.2384 \\
& CLGA1 (Cl)& $\epsilon$     & -0.3430  & -0.3400 & -0.3333 & -0.3062 & -0.3340  \\
&           & $R_{\min}/2$   &  1.9100  &  1.9941 &  1.9653 &  1.9763 &  1.9609 \\
\hline
\multirow{6}{*}{Acetone}
& CG331 (C) & $\epsilon$     & -0.0780  &-0.1017  &-0.1040  & -0.1040  &-0.1039 \\
&           & $R_{\min}/2$   &  2.0500  & 2.0068  & 2.0052  &  2.1330  & 2.0500 \\
& HGA3 (H)  & $\epsilon$     & -0.0240  &-0.0168  &-0.0196  & -0.0312  &-0.0273 \\
&           & $R_{\min}/2$   &  1.3400  & 1.4469  & 1.4596  &  1.2960  & 1.3337 \\
& OG2D3 (O) & $\epsilon$     & -0.0500  &-0.0650  &-0.0650  & -0.0650  &-0.0650 \\
&           & $R_{\min}/2$   &  1.7000  & 1.6292  & 1.5017  &  1.5233  & 1.5302 \\
\hline
\end{tabular}
}
\label{tab:lj}
\end{table}

\noindent
It is also of interest to consider changes in the LJ-parameters made
by fitting using ``Approach A", see Table \ref{tab:lj}. Atomic radii
$R_{\rm min}$ change by up to 10 \% whereas the well depths
`$\epsilon$' can change more substantially, by up to 30 \% relative to
the original CGenFF values, occasionally also reaching fitting
constraints (e.g. atom type OG2D3 for acetone). This is likely in
response to the DLNPO-MP2 level of theory used here as opposed to the
MP2 level used for CGenFF. Also, the different nature of the training
data (molecular cluster binding energies) relative to CGenFF training
data (interaction with water, condensed phase properties), requires
adjustment to implicitly account for effects such as DLNPO-MP2
dispersion, polarization and charge transfer. In contrast, the more
stable radii suggest intermolecular separation is less dependent on
the level of theory or choice of training data.\\

\begin{figure}[h!]
\centering
\includegraphics[width=1.0\textwidth]{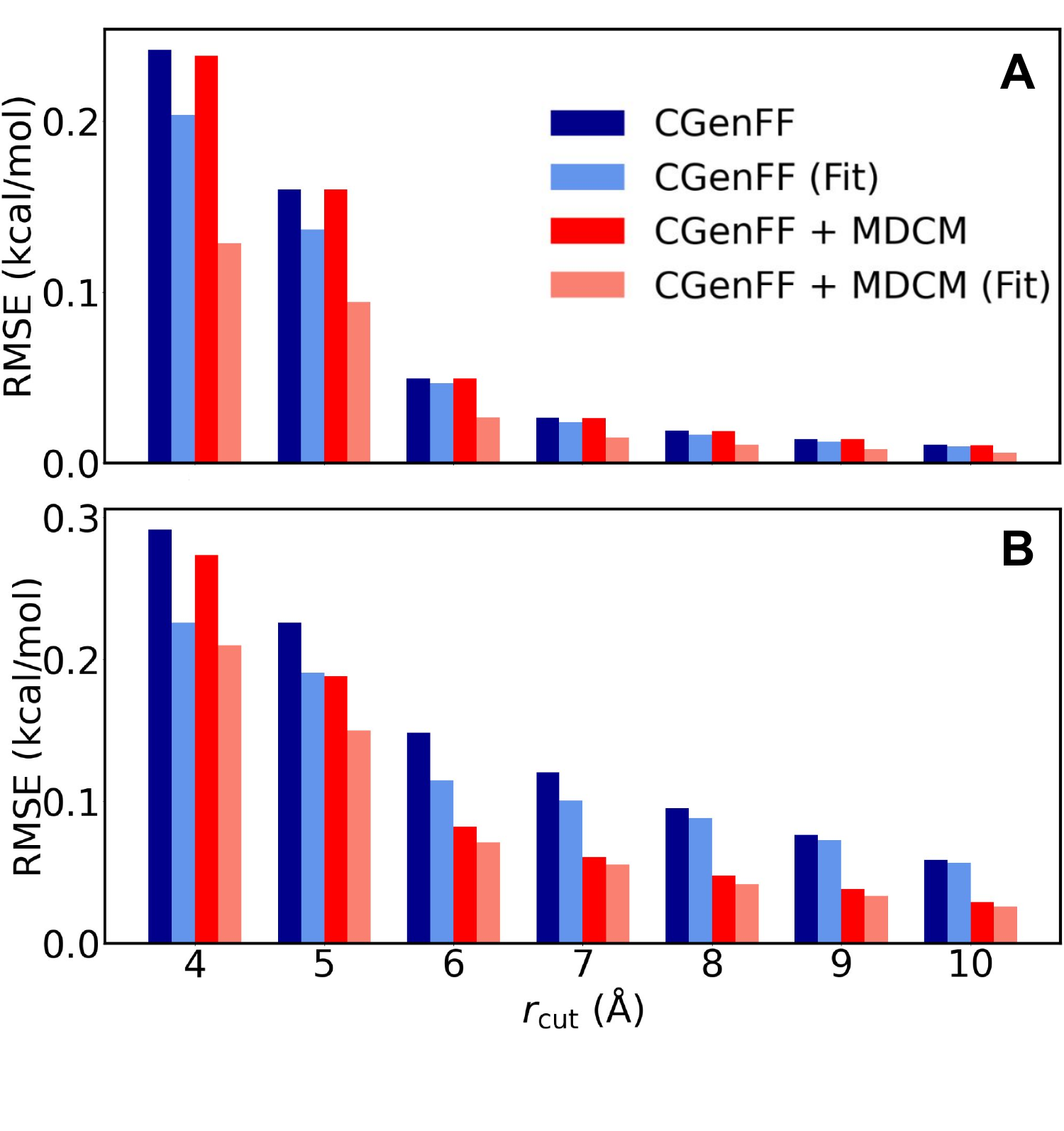}  
\caption{Performance of the unfitted dark (blue and red) and LJ-fitted
  light (blue and red) models for different values of the cutoff. Top
  panel for DCM, bottom panel for acetone. The standard deviations of
  the fits are comparable to the magnitude of the RMSE.}
\label{fig:fig5} 
\end{figure}

\noindent
One of the primary objectives of the present work is to devise and
quantitatively assess the performance of a range-separated ML/MM
energy function for which the short-range part is handled by the
ML-model and for the long-range part an empirical or modified
empirical energy function is used. Such a mixed energy function is
constructed for different values of the cut-off, defining the boundary
between short- and long-range contributions.\\

\begin{figure}[h!]
\centering
\includegraphics[width=1.0\textwidth]{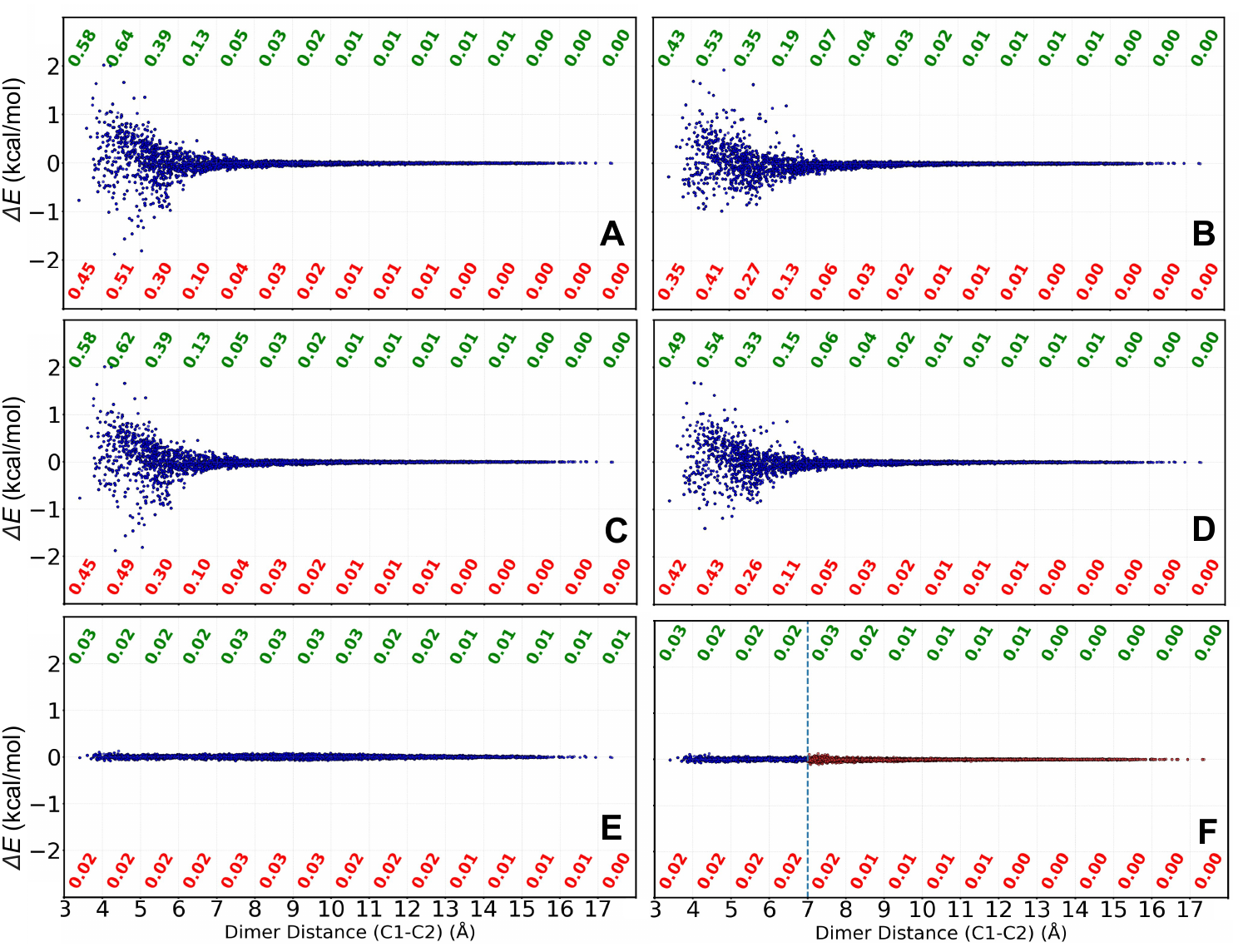}
\caption{Performance of the CGenFF (A), CGenFF+Fit (B), CGenFF+MDCM
  (C), CGenFF+MDCM+Fit (D), PhysNet (E), and Range-separated ML/MM
  with $r_{\rm cut}$=7 \AA\/ (F) models for DCM. The target quantity
  considered is $\Delta E_{j}^{\rm dimer} = E_{{\rm ref},j}^{\rm
    dimer} - E_{{\rm FF},j}^{\rm dimer}$, plotted against dimer
  separation length for the $\binom{20}{2} = \frac{20!}{2!  \cdot (20
    - 2)!} = \frac{20 \cdot 19}{2 \cdot 1} = 190$ dimers. Green and
  red numerals refer to the RMSE$(E)$ and MAE$(E)$ on the $\Delta
  E_{j}^{\rm pair}$ at every $1 {\rm \AA}$ C-C distance. Panel E for
  "PhysNet" represents $E_{j}^{\rm pair}$ of ML energy functions
  against $E_{{\rm ref},j}^{\rm pair}$. Panel F is for the hybrid
  ML/MM PES for $r\leq r_{\rm cut} = 7$ \AA\/ and LJ-fitted dimers for
  $r > r_{\rm cut}$ on the same set using "approach A" in red.}
\label{fig:fig6} 
\end{figure}

\noindent
For the present case, the ``coordinate" that was considered is the
C--C separation between two monomers. However, depending on the shape
of the monomers, other choices may be more advantageous. A
quantitative assessment for using different cutoff distances for
switching between the ML-PES and the MM contributions is shown in
Figure \ref{fig:fig5} for DCM (panel A) and acetone (panel B).\\

\noindent
The rapid improvement in all CGenFF and MDCM models, with or without
refitted LJ parameters, with increasing $r_{\rm cut}$ highlights the
advantage of handling short-range interactions with an ML
potential. Even after refitting LJ parameters to dimer reference data,
errors are significantly larger for smaller than for larger $r_{\rm
  cut}$ values for which short-range interactions are determined from
the ML-PES and only interactions beyond $r_{\rm cut}$ are handled
through the empirical energy function. For DCM the initial rapid
improvement levels off by $\sim 7$ \AA\/ for all ML/MM models. This
demonstrates that there is little improvement in accuracy beyond this
point to justify the increased complexity and computational cost of
the ML approach. For acetone the convergence with $r_{\rm cut}$ is
slower, but the trend is the same. In this case, using the more
detailed, anisotropic MDCM electrostatic model allows a significantly
shorter $r_{\rm cut}$ than the original CGenFF charges, yielding a
similar RMSE at 7 \AA\/ to CGenFF with refitted LJ at 10 \AA\/. This
highlights the advantage of using an accurate electrostatic
representation, even at separations up to 10 \AA.\\

\noindent
Figure \ref{fig:fig5} additionally shows the impact of refitting LJ
parameters as a function of $r_{\rm cut}$. With a cutoff at 4 \AA\/,
the CGenFF model improves by $\sim 15$ \% (dark vs. light blue) after
refitting LJ parameters. This effect also decreases progressively as
$r_{\rm cut}$ increases. One reason for this is that the number of
pairs that contribute to the fit progressively increases with
increasing $r_{\rm cut}$ whereas for larger $r_{\rm cut}$ the
interaction energies become less sensitive to the LJ-parameters and
the magnitude of the contributions decreases in concert. For the
CGenFF+MDCM model (red colors) the impact of fitting the LJ-parameters
is considerably more pronounced. The improvement is typically a factor
of 2 over the unfitted models and the performance of fitted
CGenFF+MDCM is uniformly better than for fitted CGenFF (compare light
blue and red in Figure \ref{fig:fig5}A). For acetone in Figure
\ref{fig:fig5}B similar observations are made although the improvement
of fitted CGenFF+MDCM over CGenFF is not as pronounced at shorter
$r_{\rm cut}$ values as for DCM.\\

\noindent
The degradation in performance of MM potentials at short range is also
highlighted explicitly in Figures \ref{fig:fig6} and \ref{sifig:fig1}
(A to D). For DCM and starting from the long-range (large $r_{\rm
  cut}$), the figures highlight the degradation in performance of
CGenFF and CGenFF+MDCM, even after refitting LJ contributions, at
separations shorter than $r_{\rm C-C} = 8$ \AA\/. In contrast, the
PhysNet PES (Figure \ref{fig:fig6}E) maintains a comparable level of
accuracy across the entire range of distances, again demonstrating the
advantage of an ML potential over a empirical FFs at short-range in
particular. Figure \ref{fig:fig6}F further illustrates benefits of
"Approach A" over ``Approach B'': for $r_{\rm cut} = 7$ \AA\/ the
errors in the ML/MM PES with ``Approach A'' are closer to PhysNet
compared with "Approach B", see Figures \ref{sifig:fig3} to
\ref{sifig:fig6}. \\

\noindent
Table \ref{tab:app-a} also shows that when applied to pairwise cluster
energies, the ML/MM models for MDCM in particular can actually exceed
the accuracy of the pure pairwise ML approach (0.42 vs 1.07 kcal/mol
for DCM, 1.08 vs 1.20 kcal/mol for acetone). By fitting to dimers
only, a small systematic shift arises in the total ML pairwise cluster
energy after summing over all 190 dimer pairs involved in each
cluster. As the MM potentials were fitted to pairwise cluster energies
rather than individual dimer energies, the systematic shift does not
arise and when combined with the short-range ML potential the overall
ML/MM accuracy exceeds that of the ML model alone.\\

\noindent
Finally, it is of interest to compare the performance of the mixed
ML/MM for two typical cutoffs ($r_{\rm cut} = 4$ and 7 \AA\/) with
PhysNet and the unfitted and LJ-fitted CGenFF and CGenFF+MDCM models,
see Figures \ref{fig:fig3} and \ref{fig:fig4}. For DCM and CGenFF
already a short cutoff of 4 \AA\/ (Figure \ref{fig:fig3}A, light blue)
brings the ML/MM into rather favourable agreement with the reference
data. Extending the cutoff to 7 \AA\/ further improves the performance
(dark blue). The same is observed for CGenFF+MDCM (panel B, light and
dark red). For acetone and CGenFF on the other hand (Figure
\ref{fig:fig4}A) a large cutoff $r_{\rm cut} = 7$ \AA\/ (dark blue) is
required for good performance. This differs for CGenFF+MDCM for which
both cutoffs perform very well and on par with each other (panel B,
light and dark red). Hence, for the more strongly interacting system
(acetone) advanced electrostatics (MDCM) provides a clear advantage
over point charge-based models such as CGenFF. Further details on the
performance of the hybrid ML/MM PESs based on different cutoffs are
reported in Tables \ref{sitab:tab1} and \ref{sitab:tab2}. \\

\subsection{$NVE$ Molecular Dynamics Simulations}
Finally, exploratory MD simulations using the mixed ML/MM-PES with
cutoff 8 \AA\/ were carried out for DCM-clusters of different
sizes. To avoid decomposition of the clusters in the gas phase a
flat-bottomed external potential was used. The simulations were
carried out in the microcanonical $NVE$ ensemble to check conservation
of total energy. This establishes correct implementation of the forces
and handling of the switching region across the ML/MM boundary.\\

\begin{figure}[H]
\centering
\includegraphics[width=1.00\textwidth]{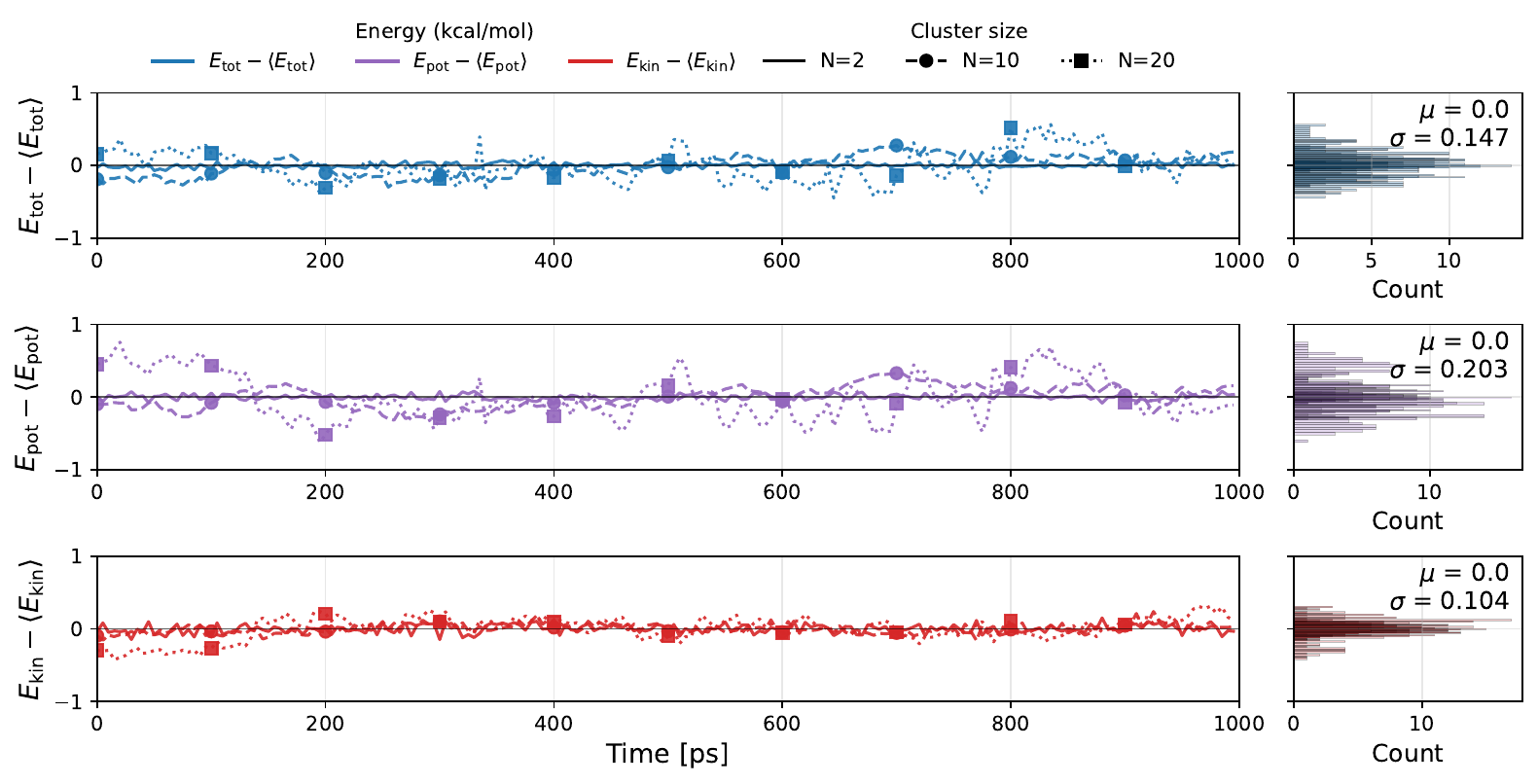}
\caption{Conservation of total energy, potential energy, and
  temperature for $NVE$ simulations ($\Delta t = 0.5$ fs) with initial
  velocities scaled to 200 K for DCM molecules ($n = 2, 10, 20$). The
  right-hand panel shows the corresponding distributions which appear
  Gaussian with no observable drift.}
\label{fig:fig7} 
\end{figure}

\noindent
As Figure \ref{fig:fig7} demonstrates, total energy is conserved on
average with no observable drift for dimers and clusters consisting of
10 and 20 DCM monomers on a time scale of 1 ns. The distributions of
total energy and its contributions are Gaussian, as expected. \\

\noindent
It is also worthwhile to briefly comment on relative timings for the
test systems considered in the present work. Using the same computer
environment (CPU) and the current implementation in pyCHARMM the
relative timings for the small model system were ${\rm MM : ML :
  ML/MM} \approx 1:25:25$. However, this ratio is likely to change for
larger systems set up together with periodic boundary conditions,
featuring a considerably larger number of pairwise interactions. The
current results suggest that such condensed-phase production
simulations will be computationally feasible even using full pairwise
ML, while savings from ML/MM splitting are predicted to be substantial
as the number of pairwise interactions increases.\\

\section{Discussion and Conclusions}
The present work presented the construction and validation of
range-separated, two-body mixed ML/MM-PESs. To illustrate and to
quantitatively assess the procedure, dichloromethane and acetone were
used. The main focus of this work concerned the feasibility and
performance of such a mixed ML/MM-PES with a particular emphasis on
narrowing down the region where to switch between the two descriptions
to a) maintain accuracy and b) improve computational performance. For
both model systems it was found that a mixed ML/MM-PES was
advantageous after refitting the LJ-parameters. Depending on the
system and MM-representation (CGenFF or CGenFF+MDCM) a shorter
($r_{\rm cut} = 4$ \AA\/) or longer ($r_{\rm cut} = 7$ \AA\/) cutoff
yielded performance comparable to that of a genuine two-body ML-PES.\\

\begin{figure}[H]
\centering \includegraphics[width=1.0\textwidth,
  height=0.38\textheight]{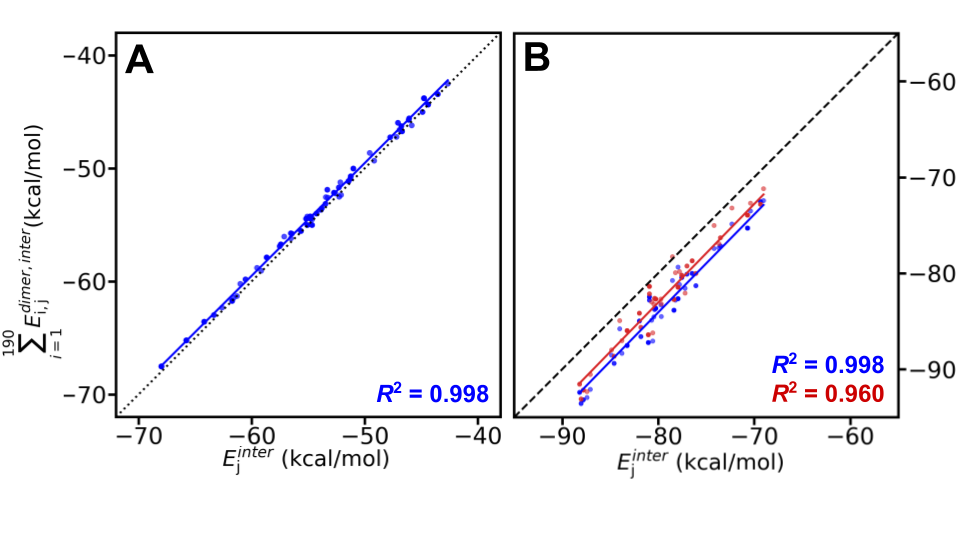}
\caption{Correlation between total cluster interaction energies
  ($x-$axis) and the sum of dimer interaction energies ($y-$axis) to
  quantify the magnitude of many-body contributions. Panels A and B
  are for DCM and Acetone, respectively. Blue symbols are for data
  based on {\it ab initio} calculations whereas red symbols use the
  trained ML-PES to evaluate the sum over all dimer energies. This
  comparison further establishes the high quality of the ML-PES. The
  solid lines are least-squares fits and the dashed lines indicate a
  1:1 correlation. For DCM, the RMSE and MAE are 0.58 and 0.50
  kcal/mol, respectively. For Acetone, the RMSE and MAE are 4.21 and
  4.05 kcal/mol (blue), and 3.23 and 3.02 kcal/mol (red),
  respectively, which is consistent with Table \ref{tab:app-a}.}
\label{fig:fig8} 
\end{figure}

\noindent
Although the quality of such models is already quite respectable (see
dark red and dark blue traces in Figures \ref{fig:fig3} and
\ref{fig:fig4}) one essential future improvement concerns inclusion of
many-body contributions. Hence, the validity of the 2-body
approximation was examined next for the DCM and acetone clusters
extracted from MD simulations (Figure \ref{fig:fig8}). In agreement
with previous work,\cite{MM.ffs:2024} the 2-body approximation
obtained by a simple sum over all dimers in each cluster agrees well
with the calculated total interaction energy evaluated by subtracting
monomer energies from the total DLPNO-MP2 cluster energy. The $R^{2}$
of 0.998 and RMSE and MAE of 0.58 and 0.50 kcal/mol confirms that
many-body effects are comparatively small for DCM in the condensed
phase. In contrast, the lower $R^{2}$ of 0.959 and higher RMSE of 4.21
kcal/mol for acetone, that is largely caused by a systematic shift in
the total interaction energy when neglecting many-body effects,
highlights that DCM is unusual in this respect, and reliable
improvement in the PES will require a many-body correction to be added
in the future.\\

\begin{table}[H]
\centering
\small
\renewcommand{\arraystretch}{1.2}
\begin{tabular}{lcccc|cccc}
\toprule
& \multicolumn{4}{c|}{\textbf{Dichloromethane (DCM)}} & \multicolumn{4}{c}{\textbf{Acetone}} \\
\cmidrule(lr){2-5}\cmidrule(lr){6-9}
\textbf{Model} & $R^2$ & RMSE & MAE & STD & $R^2$ & RMSE & MAE & STD \\
\midrule
CGenFF         & 0.92 & 3.75 & 3.08 & 2.51 & 0.78 & 4.65 & 3.95 & 2.25 \\
CGenFF + Fit   & 0.94 & 1.88 & 1.56 & 1.88 & 0.84 & 2.34 & 1.74 & 2.28 \\
\cmidrule(lr){1-9}
CGenFF+MDCM           & 0.98 & 1.61 & 1.31 & 1.54 & 0.89 & 12.01 & 11.86 & 1.90 \\
CGenFF+MDCM + Fit     & 0.98 & 1.12 & 0.90 & 1.13 & 0.87 & 2.94 & 2.41 & 1.97 \\
\cmidrule(lr){1-9}
PhysNet        & 0.99 & 1.07 & 1.01 & 0.50 & 0.96 & 3.23 & 3.02 & 1.17 \\
\bottomrule
\end{tabular}
\caption{Error statistics for DCM and acetone following ``Approach
  B". ($R^2$; RMSE/MAE/STD in kcal/mol) for correlations of cluster
  $E_{\rm Ref}$ ($x-$axis) vs.\ $E_{\rm Model}$ ($y-$axis). Left: DCM;
  see Figure~\ref{sifig:fig2}. Right: Acetone; see
  Figure~\ref{fig:fig9}.}
\label{tab:app-b}
\end{table}

\begin{figure}[htbp!]
\centering \includegraphics[width=1.00\textwidth,
  height=0.42\textheight]{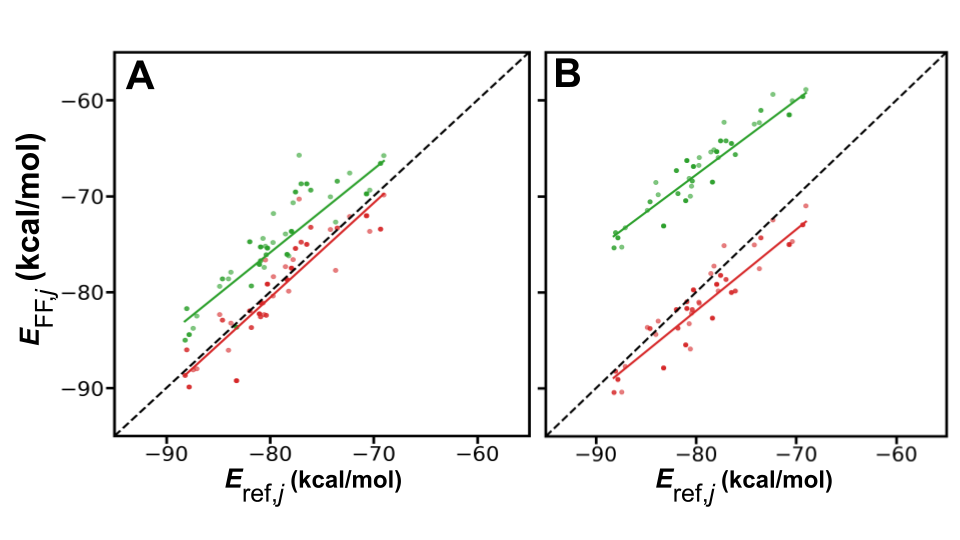}
\caption{Correlation between reference electronic structure data
  ($x-$axis) and the MM (panels A and B) energies for acetone using
  ``Approach B" for the LJ-fit. Green traces are results from using
  CGenFF, CGenFF+MDCM energies without modification.  Red traces are
  the results after refitting the LJ-parameters. For statistical
  performance measures, see Table \ref{tab:app-b}.}
\label{fig:fig9} 
\end{figure}

\noindent
{\it Approach B:} An alternative to ``Approach A" is to use total
cluster energies $E_{\rm ref}^{\rm full}$ from electronic structure
calculations as the target quantity for fitting the LJ-parameters,
implicitly adding an averaged many-body correction to LJ terms at the
expense of accuracy in the two-body description. While parameters from
``Approach A'' can be corrected by adding an explicit many-body term,
making the resulting ML/MM potential highly transferable, the
implicit, averaged many-body correction of ``Approach B'' is more
approximate and valid only for the system for which it was fitted.\\

\noindent
To test ``Approach B'' for DCM and acetone, total cluster energies
including many-body interactions beyond the two-body approximation
were used. Figures \ref{fig:fig9} and \ref{sifig:fig2} compare the
cluster energies from electronic structure calculations for acetone
with those from the CGenFF (panel A) and the CGenFF+MDCM (panel B)
models (green symbols and lines). After readjusting the LJ-parameters
(Table \ref{tab:lj}) to best describe the cluster energies both models
manifestly improve (red symbols and lines): the [RMSE$(E)$, and
  MAE$(E)$] improve from [4.65, 3.95] kcal/mol to [2.34, 1.74]
kcal/mol for CGenFF and from [12.01, 11.86] kcal/mol to [2.94, 2.41]
kcal/mol. For DCM the improvements are even larger: from [3.75, 3.08]
kcal/mol to [1.88, 1.56] kcal/mol for CGenFF and from [1.61, 1.31]
kcal/mol to [1.12, 0.90] kcal/mol, see Table \ref{tab:app-b}. Again
the improvement arises predominantly from shifting the mean of the FF
distribution towards the reference, removing the systematic error,
while the width of the distribution, representing the relative cluster
energies and quantified using the standard deviation, is largely
unchanged. As a result, the DCM 2-body ML description is closer to the
total cluster energy (RMSE 1.88 kcal/mol CGenFF+Fit, 1.12 kcal/mol
MDCM+Fit vs. 1.01 kcal/mol PhysNet 2-body), despite neglecting
many-body effects, while the acetone 2-body ML description (2.34
kcal/mol CGenFF+Fit, 2.94 MDCM+Fit vs. 3.23 kcal/mol PhysNet 2-body)
has a slightly larger systematic shift (overestimated formation
energies) than the FFs but still a smaller standard deviation (better
relative energies) despite neglecting many-body effects entirely.\\

\noindent
Changes in the LJ-parameters following ``Approach B'' are also
reported in Table \ref{tab:lj}. While parameters obtained are
comparable to those of ``Approach A'', the changes again highlight
that $\epsilon$ well-depth parameters are more sensitive to the type
and level of theory of reference data than the atom radii.\\

\noindent
It is useful to compare these findings with previous work primarily
carried out for water using the MB-pol and q-AQUA
models.\cite{paesani:2013,paesani:2014,paesani:2014-2,paesani:2024,bowman:2022}
The two water models are on-par when compared with CCSD(T) reference
energies. One relevant benchmark to consider is the range of two-,
three-, and four-body energies in the reference data. For the q-AQUA
model they cover 300 kcal/mol, 20 kcal/mol, and 3 kcal/mol,
respectively. In other words, the magnitude of the contributions
decrease by one order of magnitude when increasing the order by
one. On the other hand, for particular applications, the importance of
the higher many-body contributions relative to the two-body energy can
deviate from this. For water hexamer cluster energies the ratio of
2:3:4-body contributions is closer to 1:0.3:0.03. In other words:
three-body contributions are considerably more important for water
hexamers than for random structures which are part of the training
set. It should, however, be noted that for water 3-body contributions
are expected to be particularly relevant and for the two systems
discussed in the present work - dichloromethane and acetone - their
importance is expected to be smaller.\\

\noindent
In summary, the present work introduces a practical approach for
2-body, range-separated mixed ML/MM-PES for condensed systems. The
range-separated ML/MM approach was validated for dichloromethane and
acetone and showed accuracies for cluster energies on par with a
ML-PES trained on monomers and dimers with considerably fewer pairwise
interactions handled through explicit ML energies. Using CGenFF and
CGenFF+MDCM as the MM description in the LJ-fits, the improvement of
the ML/MM-PES levels off by $\sim 7$ \AA\/ for DCM. For acetone, the
anisotropic MDCM electrostatic model also allows a shorter $\sim 7$
\AA\/ cutoff whereas with the original CGenFF energy function only
saturates by $\sim 10$ \AA\/. This highlights the advantage of using
an accurate electrostatic representation for such an
ML/MM-PES. Importantly, the present approach can also be extended to
mixed chemical systems and systems containing ions. Future extensions
concern the inclusion of many-body contributions.\\

\section*{Supporting Information}
Figures reporting the performance for fitted models depending on the
cutoff for both systems and Tables with RMSE($E$) and standard
deviations for the LJ-fitted values.

\section*{Data Availability}
The data for the present study are available from
\url{https://github.com/MMunibas/mlmm} upon publication.

\section*{Acknowledgment}
We thank the Swiss National Science Foundation (grants
200020{\_}219779 and 200021{\_}215088) (to MM), and the University of
Basel for supporting this work. \\

\clearpage
\bibliography{refs}

\clearpage

\renewcommand{\thetable}{S\arabic{table}}
\renewcommand{\thefigure}{S\arabic{figure}}
\renewcommand{\thesection}{S\arabic{section}}
\renewcommand{\d}{\text{d}}
\setcounter{figure}{0}  
\setcounter{section}{0}  
\setcounter{table}{0}

\newpage

\noindent
{\bf SUPPORTING INFORMATION: Explicit, Machine-Learned Two-Body
  Potentials for Molecular Simulations}

\section{Switching Function}

\begin{figure}[H]
    \centering
    \includegraphics[width=0.5\linewidth]{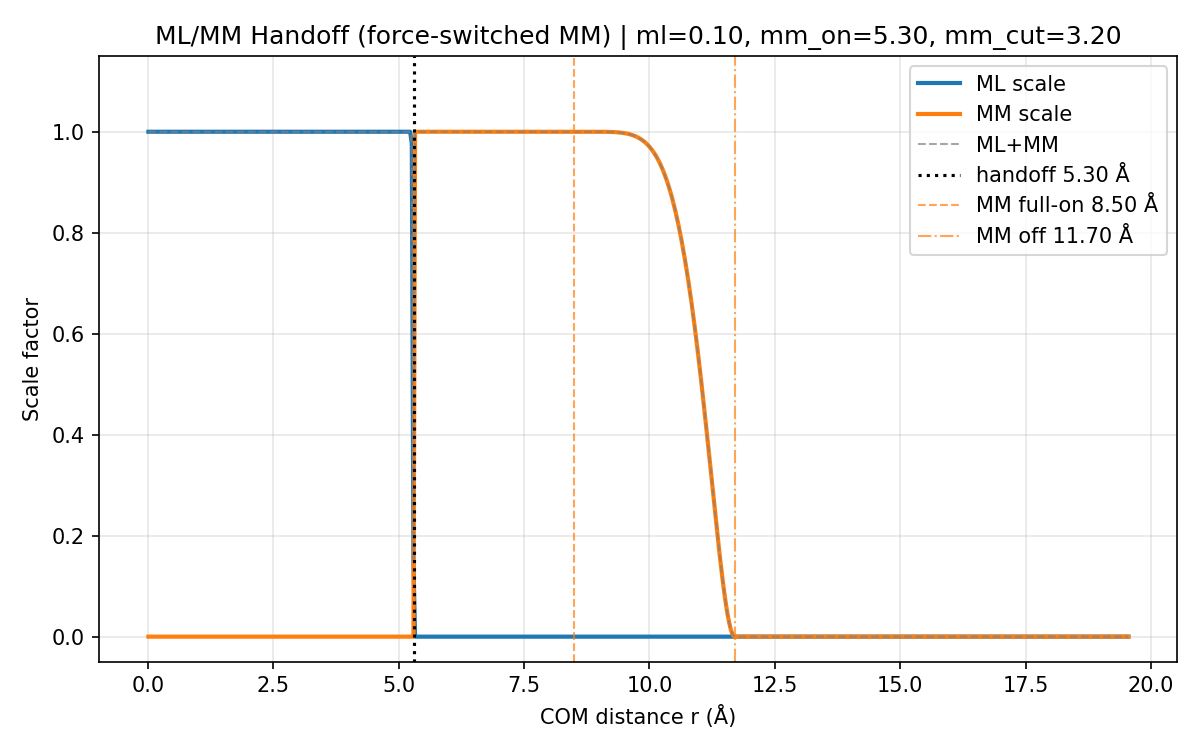}
    \includegraphics[width=0.5\linewidth]{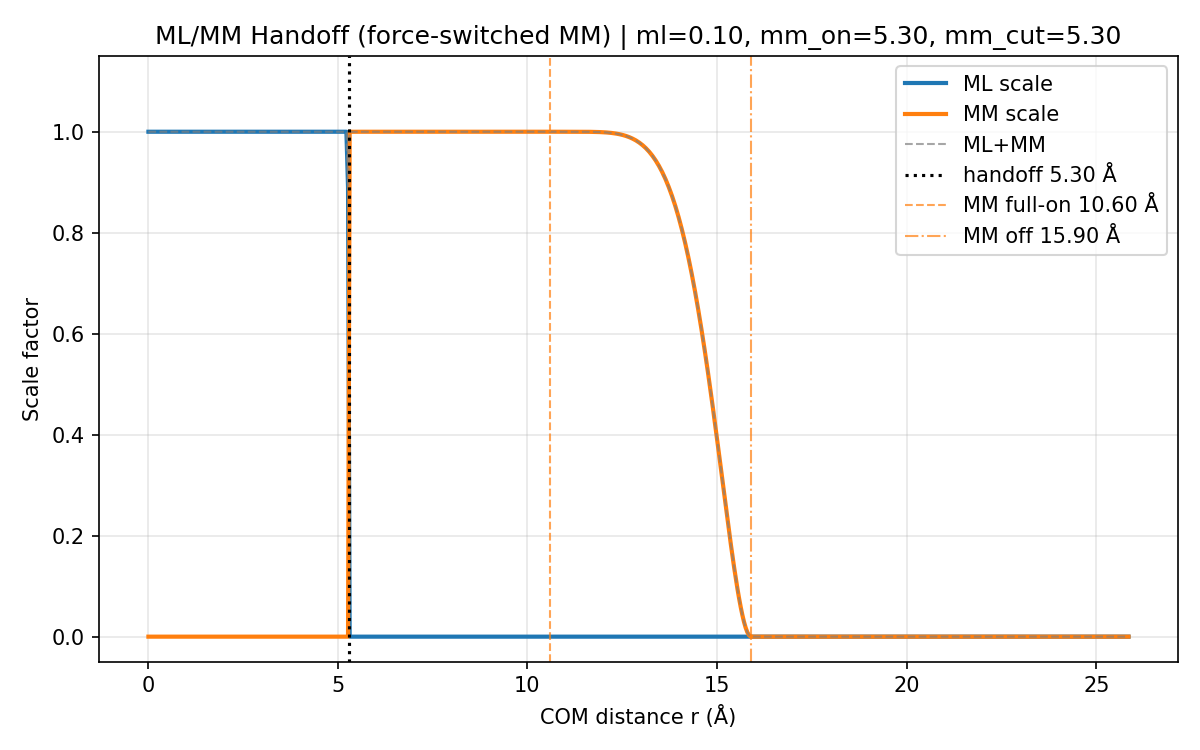}
    \includegraphics[width=0.5\linewidth]{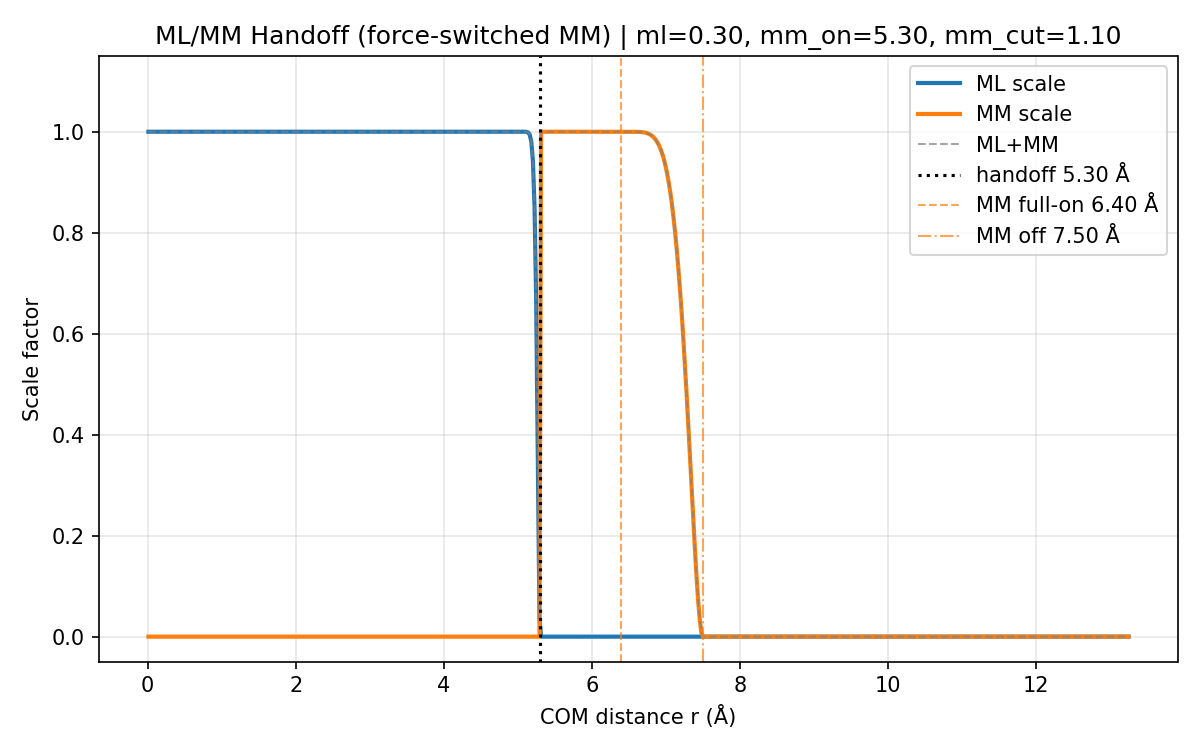}
    \includegraphics[width=0.5\linewidth]{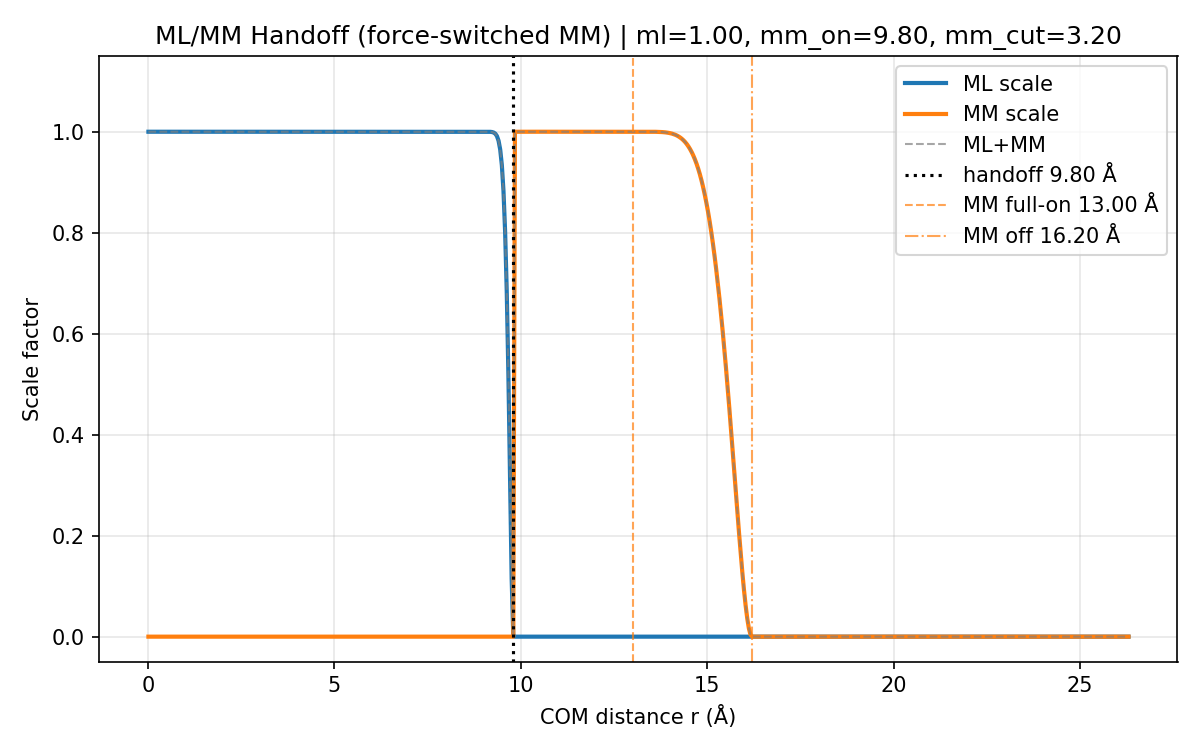}
    \caption{Examples of the cut-offs and switching functions used.}
    \label{sifig:fig7}
\end{figure}

\section{Dimer distance vs Dimer Energy Error}

\begin{figure}[H]
\centering
\includegraphics[width=1.05\textwidth, height=0.75\textheight]{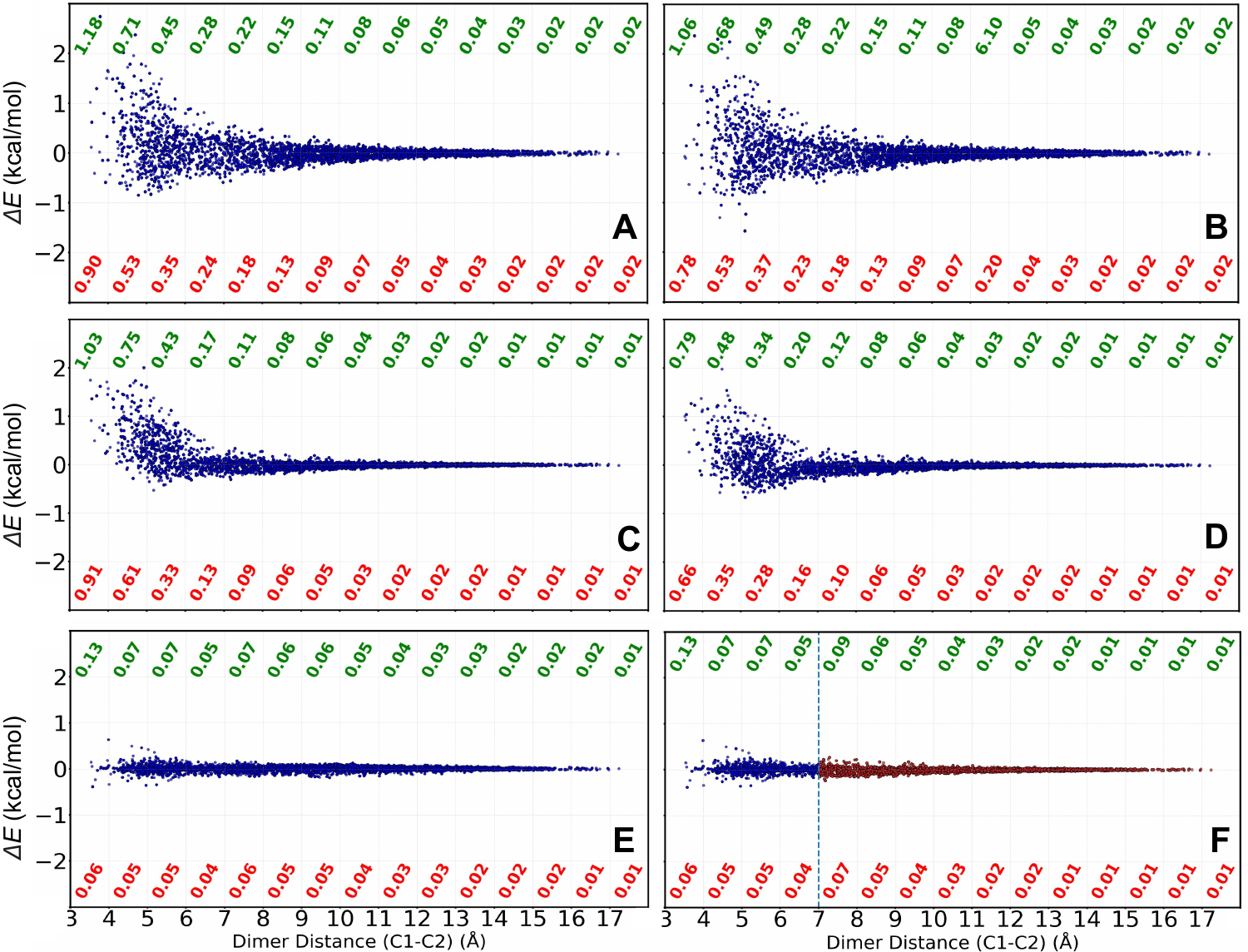}  
\caption{$\Delta E_{j}^{\rm pair} = E_{{\rm ref},j}^{\rm pair} -
  E_{{\rm FF},j}^{\rm pair}$ for acetone, plotted against dimer
  separation length. A total of $\binom{20}{2} = \frac{20!}{2! \cdot
    (20 - 2)!} = \frac{20 \cdot 19}{2 \cdot 1} = 190$ dimers are
  analyzed. Panels A to D: CGenFF with fitted LJ, CGenFF+MDCM, and
  CGenFF+MDCM with fitted LJ using ``Approach B''. Panel E: the
  PhysNet model $E_{j}^{\rm pair}$ of ML energy functions against
  $E_{{\rm ref},j}^{\rm pair}$. Panel F: PhysNet for $r\leq r_{\rm
    cut}$ and CGenFF+MDCM with fitted LJ dimers for $r\geq r_{\rm
    cut}$ using "approach A".}
\label{sifig:fig1} 
\end{figure}

\begin{figure}[H]
\centering
\includegraphics[width=1.0\textwidth]{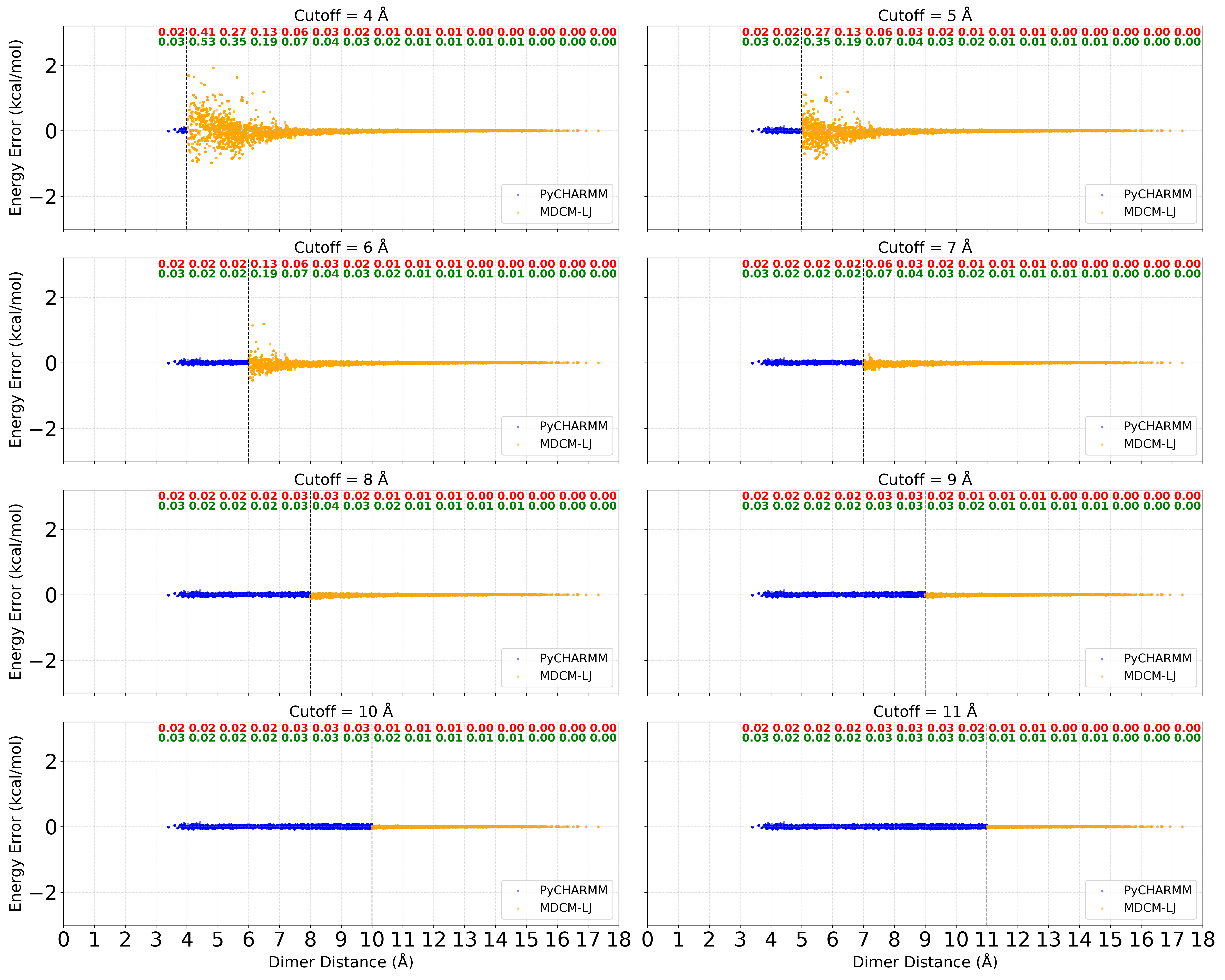}
\caption{\textbf{DCM:Model=CGenFF+LJ}.  $\Delta E_{\text{Dimer}} =
\begin{cases}
E_{\text{PhysNet}} - E_{\text{ORCA}}, & \text{if } r < r_{\text{cut}} \\
E_{\text{Model}} - E_{\text{ORCA}}, & \text{if } r \geq r_{\text{cut}}
\end{cases}$. }
\label{sifig:fig3} 
\end{figure}

\begin{figure}[H]
\centering
\includegraphics[width=1.0\textwidth]{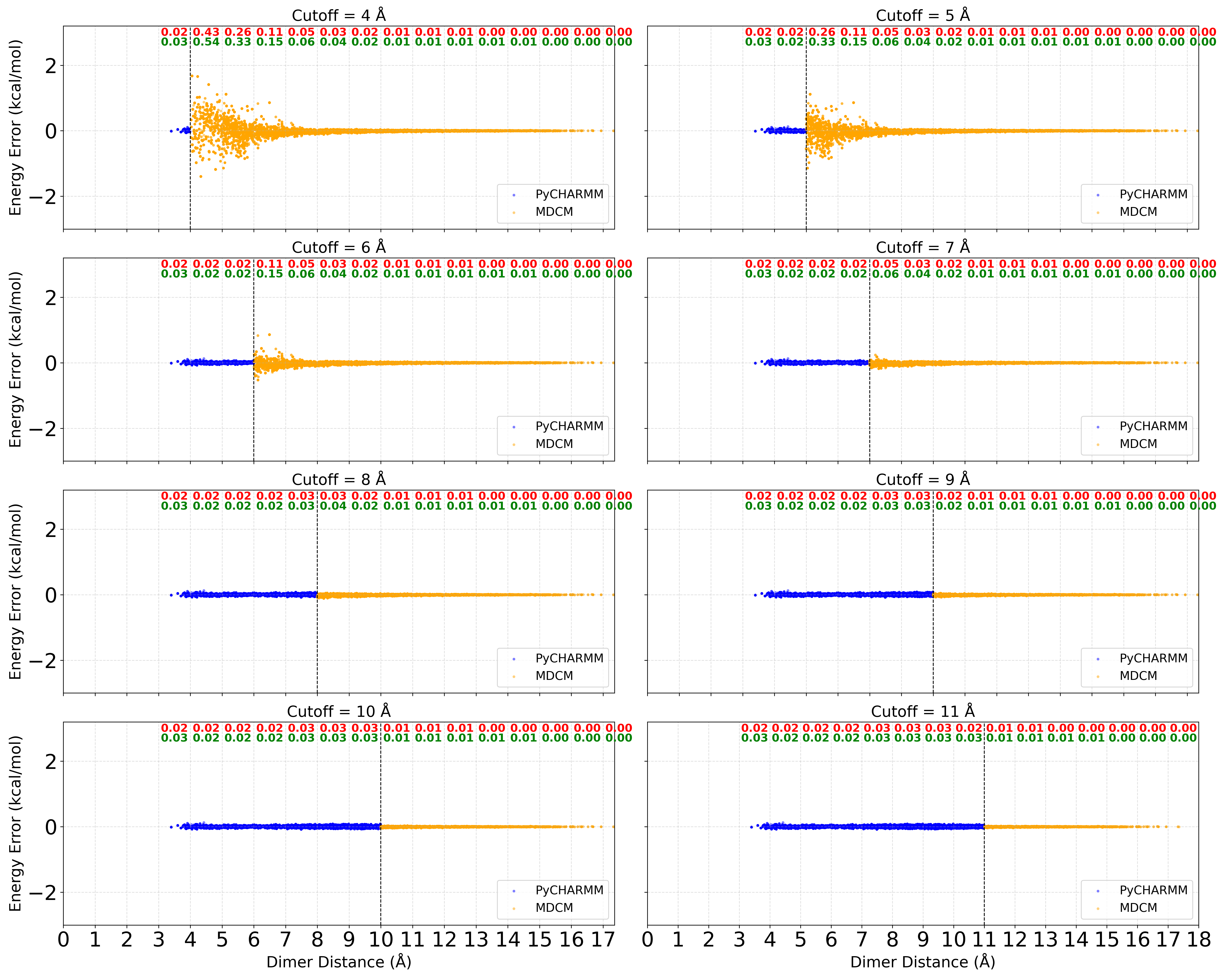}  
\caption{\textbf{DCM: Model = MDCM+LJ}.
$\Delta E_{\text{Dimer}} = 
\begin{cases}
E_{\text{PhysNet}} - E_{\text{ORCA}}, & \text{if } r < r_{\text{cut}} \\
E_{\text{Model}} - E_{\text{ORCA}}, & \text{if } r \geq r_{\text{cut}}
\end{cases}$. }
\label{sifig:fig4} 
\end{figure}

\subsection{Acetone}

\begin{figure}[H]
\centering
\includegraphics[width=1.0\textwidth]{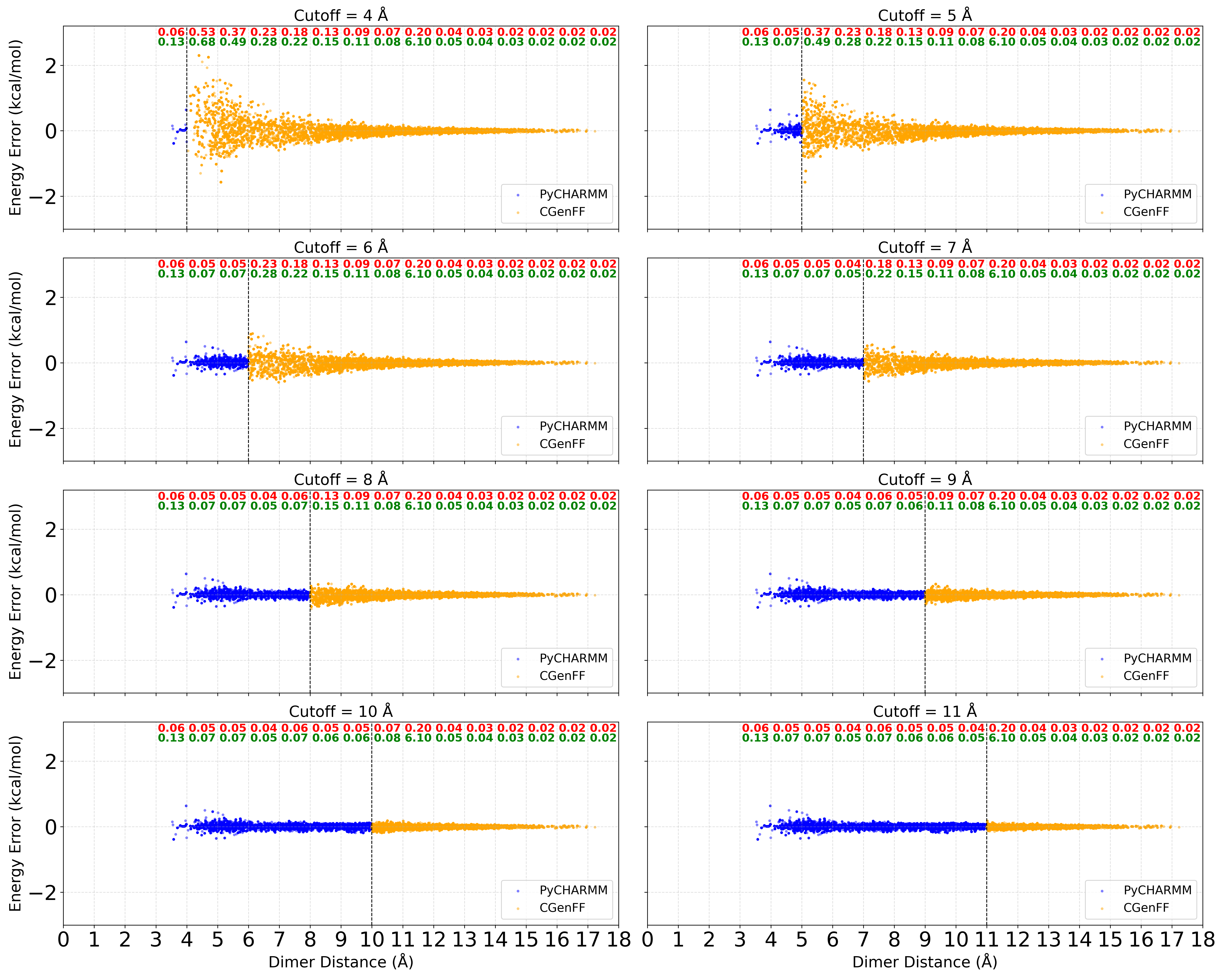}
\caption{\textbf{Acetone: Model = CGenFF+LJ}.  $\Delta
  E_{\text{Dimer}} =
\begin{cases}
E_{\text{PhysNet}} - E_{\text{ORCA}}, & \text{if } r < r_{\text{cut}} \\
E_{\text{Model}} - E_{\text{ORCA}}, & \text{if } r \geq r_{\text{cut}}
\end{cases}$.}
\label{sifig:fig5} 
\end{figure}

\begin{figure}[H]
\centering
\includegraphics[width=1.0\textwidth]{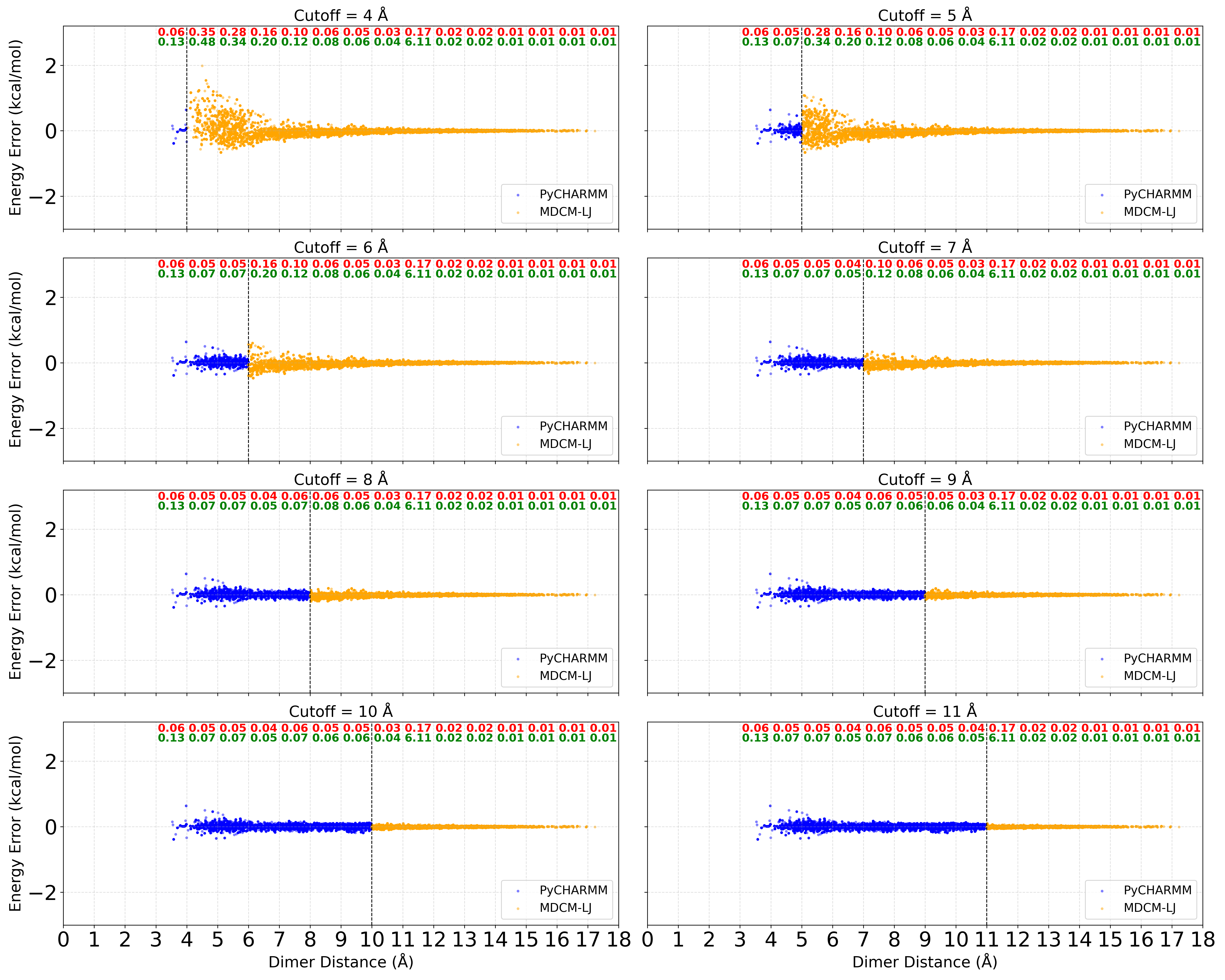}
\caption{\textbf{Acetone:Model=MDCM+LJ}.
$\Delta E_{\text{Dimer}} = 
\begin{cases}
E_{\text{PhysNet}} - E_{\text{ORCA}}, & \text{if } r < r_{\text{cut}}
\\ E_{\text{Model}} - E_{\text{ORCA}}, & \text{if } r \geq
r_{\text{cut}}
\end{cases}$.}
\label{sifig:fig6} 
\end{figure}

\begin{figure}[htbp!]
\centering \includegraphics[width=1.00\textwidth,
  height=0.42\textheight]{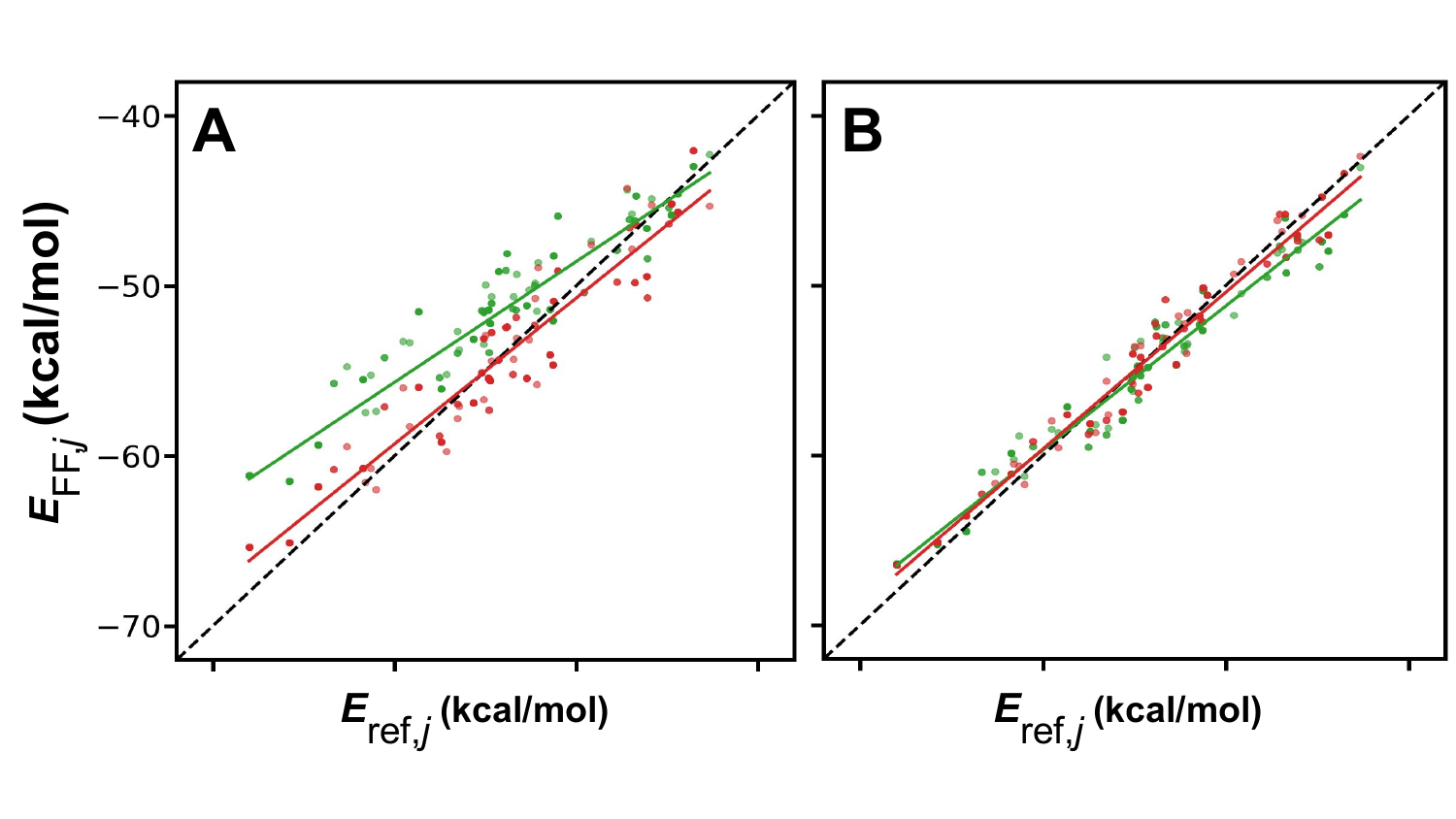}
\caption{Correlation between reference electronic structure data
  ($x-$axis) and the MM (panels A and B) energies for DCM using
  ``Approach B" for the LJ-fit. Green traces are results from using
  CGenFF, CGenFF+MDCM energies without modification.  Red traces are
  the results after refitting the LJ-parameters. For statistical
  performance measures, see Table \ref{tab:app-b}.}
\label{sifig:fig2} 
\end{figure}

\newpage
\section{LJ-Fits: Tables}

\begin{table}[h]
\centering
\caption{Statistical performance for the ML/MM PES for DCM following
  ``Approach A''. The table reports RMSE and MAE in (kcal/mol) for the
  dimer interaction energies evaluated for distances $r <
  r_\mathrm{cut}$.  For each value of $r_\mathrm{cut}$, the RMSE and
  MAE are reported in the first and second row, respectively.}
\label{sitab:tab3}
\begin{tabular}{c c cc cc c c c}
\toprule
Cutoff & Metric 
& \multicolumn{2}{c}{CGenFF} 
& \multicolumn{2}{c}{MDCM} 
& $N_\mathrm{total}$ & $N_{<\mathrm{cut}}$ & $N_{\ge\mathrm{cut}}$ \\
\cmidrule(lr){3-4} \cmidrule(lr){5-6}
(\AA) & 
& No fit & Fit & No fit & Fit &  &  &  \\
\midrule

4  & RMSE & 0.2417 & 0.2036 & 0.2383 & 0.1286 & 37949 & 404   & 37545 \\
   & MAE  & 0.2414 & 0.2029 & 0.2380 & 0.1244 &       &       &       \\

5  & RMSE & 0.1601 & 0.1369 & 0.1601 & 0.09426 & 37949 & 3637  & 34312 \\
   & MAE  & 0.1598 & 0.1369 & 0.1598 & 0.09424 &      &       &       \\

6  & RMSE & 0.0498 & 0.04671 & 0.04966 & 0.02687 & 37949 & 8979  & 28970 \\
   & MAE  & 0.04894 & 0.04652 & 0.04879 & 0.02642 &       &       &       \\

7  & RMSE & 0.02652 & 0.02391 & 0.02638 & 0.01497 & 37949 & 12263 & 25686 \\
   & MAE  & 0.02546 & 0.02340 & 0.02532 & 0.01407 &       &       &       \\

8  & RMSE & 0.01895 & 0.01660 & 0.01886 & 0.01077 & 37949 & 16130 & 21819 \\
   & MAE  & 0.01799 & 0.01658 & 0.01789 & 0.01058 &       &       &       \\

9  & RMSE & 0.01436 & 0.01277 & 0.01426 & 0.008333 & 37949 & 20335 & 17614 \\
   & MAE  & 0.01366 & 0.01276 & 0.01355 & 0.008177 &       &       &       \\

10 & RMSE & 0.01083 & 0.009899 & 0.01073 & 0.00623 & 37949 & 25360 & 12589 \\
   & MAE  & 0.01037 & 0.009885 & 0.01026 & 0.006219 &       &       &       \\
\bottomrule
\end{tabular}
\end{table}

\begin{table}[h]
\centering
\caption{Statistical performance for the ML/MM PES for acetone
  following ``Approach A''. The table reports RMSE and MAE in
  (kcal/mol) for the dimer interaction energies evaluated for
  distances $r < r_\mathrm{cut}$.  For each value of $r_\mathrm{cut}$,
  the RMSE and MAE are reported in the first and second row,
  respectively.}
\label{sitab:tab4}
\begin{tabular}{c c cc cc c c c}
\toprule Cutoff & Metric & \multicolumn{2}{c}{CGenFF} &
\multicolumn{2}{c}{MDCM} & $N_\mathrm{total}$ & $N_{<\mathrm{cut}}$ &
$N_{\ge\mathrm{cut}}$ \\ \cmidrule(lr){3-4} \cmidrule(lr){5-6} (\AA) &
& No fit & Fit & No fit & Fit & & & \\ \midrule

4  & RMSE & 0.2914 & 0.2261 & 0.2738 & 0.2103 & 18999 & 112   & 18887 \\
   & MAE  & 0.2889 & 0.2462 & 0.2621 & 0.2523 &       &       &       \\

5  & RMSE & 0.2260 & 0.1910 & 0.1884 & 0.1502 & 18999 & 1536  & 17463 \\
   & MAE  & 0.2257 & 0.2118 & 0.1848 & 0.1764 &       &       &       \\

6  & RMSE & 0.1484 & 0.1151 & 0.08228 & 0.07124 & 18999 & 4333  & 14666 \\
   & MAE  & 0.1483 & 0.1318 & 0.08195 & 0.07887 &       &       &       \\

7  & RMSE & 0.1208 & 0.1008 & 0.06082 & 0.05571 & 18999 & 6027  & 12972 \\
   & MAE  & 0.1202 & 0.1181 & 0.05976 & 0.05615 &       &       &       \\

8  & RMSE & 0.09542 & 0.08842 & 0.04791 & 0.04195 & 18999 & 7783  & 11216 \\
   & MAE  & 0.09472 & 0.09212 & 0.04696 & 0.04391 &       &       &       \\

9  & RMSE & 0.07667 & 0.07301 & 0.03841 & 0.03352 & 18999 & 9929  & 9070 \\
   & MAE  & 0.07650 & 0.07580 & 0.03809 & 0.03591 &       &       &       \\

10 & RMSE & 0.05909 & 0.05719 & 0.02933 & 0.02599 & 18999 & 12379 & 6620 \\
   & MAE  & 0.05904 & 0.07841 & 0.02920 & 0.02732 &       &       &       \\
\bottomrule
\end{tabular}
\end{table}

\begin{table}[h]
    \centering
    \small
    \renewcommand{\arraystretch}{1.2}
    \begin{tabular}{c|cc|cc|cc|cc}
        \toprule
        $r_{\rm cut}$ (\AA) & \multicolumn{2}{c|}{\textbf{CGenFF}} & \multicolumn{2}{c|}{\textbf{CGenFF LJ}} & \multicolumn{2}{c|}{\textbf{MDCM}} & \multicolumn{2}{c}{\textbf{MDCM LJ}} \\
        \cmidrule(lr){2-3} \cmidrule(lr){4-5} \cmidrule(lr){6-7} \cmidrule(lr){8-9}
        & RMSE & STD & RMSE & STD & RMSE & STD & RMSE & STD \\
        \midrule
\multirow{2}{*}{4}   & 3.54  & 2.47 & 1.73(1.75) & 1.729(1.67) & 3.83 & 2.72 & 1.89(1.76) & 1.83 1.51)\\
                     & 3.26  & 2.53 & 1.84(1.70) & 1.772(1.70) & 3.56 & 2.77 & 1.89(1.52) & 1.89(1.49)\\
\cline{2-9}

\multirow{2}{*}{5}   & 2.30 & 2.00 & 3.31(1.54) & 1.534(1.44) & 2.30 & 2.002 & 2.62(1.34) & 1.56(1.11)\\
                     & 2.60 & 2.04 & 3.75(1.49) & 1.584(1.49) & 2.60 & 2.06  & 3.04(1.38) & 1.61(1.33)\\
\cline{2-9}

\multirow{2}{*}{6}   & 0.84 & 0.56 & 2.42(0.48) & 0.992(0.48) & 0.85 & 0.558 & 1.52(0.40) & 0.75(0.40)\\
                     & 1.19 & 0.43 & 2.82(0.53) & 0.874(0.37) & 1.19 & 0.439 & 1.90(0.55) & 0.62(0.49)\\
\cline{2-9}

\multirow{2}{*}{7}   & 0.44 & 0.41 & 1.25(0.46) & 0.383(0.38) & 0.43 & 0.407 & 0.76(0.41) & 0.39(0.30)\\
                     & 0.67 & 0.27 & 1.69(0.40) & 0.282(0.26) & 0.67 & 0.259 & 1.15(0.42) & 0.26(0.27)\\
\cline{2-9}

\multirow{2}{*}{8}   & 0.51 & 0.43 & 0.48(0.64) & 0.416(0.43) & 0.50 & 0.426 & 0.42(0.79) & 0.42(0.36)\\
                     & 0.36 & 0.30 & 0.78(0.29) & 0.314(0.29) & 0.36 & 0.296 & 0.55(0.33) & 0.30(0.28)\\
\cline{2-9}

\multirow{2}{*}{9}   & 0.78 & 0.41 & 0.55(0.74) & 0.397(0.41) & 0.77 & 0.409 & 0.66(0.63) & 0.40(0.52)\\
                     & 0.32 & 0.26 & 0.27(0.29) & 0.253(0.23) & 0.31 & 0.255 & 0.26(0.25) & 0.25(0.22)\\
\cline{2-9}

\multirow{2}{*}{10}  & 1.15 & 0.49 & 1.05(1.13) & 0.484(0.48) & 1.14 & 0.48 & 1.10(1.10) & 0.48(0.47)\\
                     & 0.66 & 0.37 & 0.57(0.65) & 0.341(0.56) & 0.66 & 0.33 & 0.61(0.62) & 0.34(0.34)\\

        \bottomrule
    \end{tabular}
\caption{\emph{Dichloromethane.} Comparison of RMSE and standard
  deviation (STD) values for different ML/MM hybrid models. The
  cumulative cluster representation (Model) is constructed such that
  interactions with $r_i \le r_{\rm cut}$ are described by PhysNet,
  whereas interactions with $r_i \ge r_{\rm cut}$ are described by
  either CGenFF or MDCM and their corresponding LJ-refitted variants.
  Here $r_i$ denotes the C--C distance. Each cutoff $r_{\rm cut}$ is
  represented by two rows separated by horizontal lines. Inside the
  row \emph{Top:} $\sum_{i=1}^{190}
  E^{\mathrm{dimer,inter}}_{i}(\mathrm{Model})$
  vs.\ $E_{\mathrm{cluster}}^{\mathrm{ref}}$, benchmarked against pure
  PhysNet energies, with RMSE 1.572 kcal/mol and STD 0.501 kcal/mol
  over the entire cutoff.  \emph{Bottom:} $\sum_{i=1}^{190}
  E^{\mathrm{dimer,inter}}_{i}(\mathrm{Model})$ vs.\ $\sum_{i=1}^{190}
  E^{\mathrm{dimer,inter}}_{i,\mathrm{ref}}$, with RMSE 1.071 kcal/mol
  and STD 0.341. Numbers in parentheses correspond to metrics obtained
  from $r_{\rm cut}$-dependent LJ parameter optimization using
  ``Approach A'' (See Figure \ref{fig:fig5}). }
    \label{sitab:tab1}
\end{table}

\begin{table}[h]
    \centering
    \small
    \renewcommand{\arraystretch}{1.2}
    \begin{tabular}{c|cc|cc|cc|cc}
        \toprule
        $r_{\rm cut}$ (\AA) & \multicolumn{2}{c|}{\textbf{CGenFF}} & \multicolumn{2}{c|}{\textbf{CGenFF L-J}} & \multicolumn{2}{c|}{\textbf{MDCM}} & \multicolumn{2}{c}{\textbf{MDCM L-J}}\\
        \cmidrule(lr){2-3} \cmidrule(lr){4-5} \cmidrule(lr){6-7} \cmidrule(lr){8-9}
        & RMSE & STD & RMSE & STD & RMSE & STD & RMSE & STD\\
        \midrule
        \multirow{2}{*}{4}  & 6.64 & 6.16 & 2.73(2.82) & 2.44(2.39) & 11.03 & 1.96 & 3.47(6.41) & 1.99(2.12)\\
                            & 8.99 & 6.17 & 3.82(6.04) & 2.58(2.30) & 15.00 & 1.67 & 2.13(2.79) & 1.76(1.96)\\
        \cline{2-9}
        \multirow{2}{*}{5}  & 6.56 & 6.04 & 5.32(4.06) & 2.21(2.10) & 2.93 & 1.63 & 6.77(10.01) & 1.46(1.54)\\
                            & 6.44 & 6.26 & 2.66(2.57) & 2.54(2.50) & 6.80 & 2.08 & 3.06(6.08) & 1.67(1.67)\\
        \cline{2-9}
        \multirow{2}{*}{6}  & 7.85 & 5.92 & 5.96(5.61) & 1.67(1.68) & 4.98 & 1.32 & 7.95(9.52) & 1.32(1.29)\\
                            & 6.11 & 6.01 & 2.48(2.23) & 1.84(1.81) & 1.34 & 1.11 & 3.94(5.50) & 1.11(1.15)\\
        \cline{2-9}
        \multirow{2}{*}{7}  & 8.29 & 5.81 & 6.22(6.05) & 1.59(1.59) & 5.32 & 1.26 & 6.53(7.36) & 1.25(1.27)\\
                            & 6.13 & 5.84 & 2.39(1.53) & 1.36(1.37) & 1.46 & 0.93 & 2.53(1.08) & 0.94(0.96)\\
        \cline{2-9}
        \multirow{2}{*}{8}  & 8.04 & 5.79 & 5.54(5.46) & 1.49(1.50) & 4.93 & 1.27 & 5.54(5.96) & 1.25(1.27)\\
                            & 5.99 & 5.80 & 1.72(1.67) & 1.15(1.15) & 1.10 & 0.84 & 1.57(0.76) & 0.82(0.83)\\
        \cline{2-9}
        \multirow{2}{*}{9}  & 7.30 & 5.71 & 4.38(4.34) & 1.40(1.40) & 4.16 & 1.29 & 4.48(4.96) & 1.28(1.29)\\
                            & 5.73 & 5.71 & 0.98(0.98) & 0.98(0.98) & 0.79 & 0.78 & 0.80(0.64) & 0.77(0.58)\\
        \cline{2-9}
        \multirow{2}{*}{10} & 6.94 & 5.71 & 3.74(3.73) & 1.34(1.34) & 3.66 & 1.26 & 3.80(3.89) & 1.24(1.25)\\
                            & 5.67 & 5.67 & 1.05(1.06) & 0.89(0.89) & 0.97 & 0.75 & 0.87(0.73) & 0.74(0.74)\\
        \bottomrule
    \end{tabular}
\caption{\emph{Acetone.} Comparison of RMSE and standard deviation
  (STD) values for different ML/MM hybrid models. The cumulative
  cluster representation (Model) is constructed such that interactions
  with $r_i \le r_{\rm cut}$ are described by PhysNet, whereas
  interactions with $r_i \ge r_{\rm cut}$ are described by either
  CGenFF or MDCM and their corresponding LJ-refitted variants.  Here
  $r_i$ denotes the carbonyl C--C distance. Each cutoff $r_{\rm cut}$
  is represented by two rows separated by horizontal lines. Inside the
  row \emph{Top:} $\sum_{i=1}^{190}
  E^{\mathrm{dimer,inter}}_{i}(\mathrm{Model})$
  vs.\ $E_{\mathrm{cluster}}^{\mathrm{ref}}$, benchmarked against pure
  PhysNet energies, with RMSE = 3.230 kcal/mol and STD = 1.167
  kcal/mol over the entire cutoff.  \emph{Bottom:} $\sum_{i=1}^{190}
  E^{\mathrm{dimer,inter}}_{i}(\mathrm{Model})$ vs.\ $\sum_{i=1}^{190}
  E^{\mathrm{dimer,inter}}_{i,\mathrm{ref}}$, with RMSE = 1.195
  kcal/mol and STD = 0.589 kcal/mol. Numbers in parentheses correspond
  to metrics obtained from $r_{\rm cut}$-dependent LJ parameter
  optimization using ``Approach A'' (See Figure \ref{fig:fig5}).}
    \label{sitab:tab2}
\end{table}

\end{document}